\documentclass[%
superscriptaddress,
 amsmath,amssymb,
 aps,
 twocolumn,
 nobalancelastpage,
 raggedbottom,
 longbibliography
]{revtex4-1}
\usepackage{graphicx}
\usepackage{bm}
\usepackage{braket}
\usepackage{subfiles}
\usepackage{array}
\usepackage{float}
\usepackage{multirow}
\usepackage{amsmath}
\usepackage{amssymb}
\usepackage{graphicx}
\PassOptionsToPackage{normalem}{ulem}
\usepackage{ulem}
\usepackage{qcircuit}
\usepackage{xcolor}
\usepackage{siunitx}

\makeatletter
\floatstyle{ruled}
\newfloat{algorithm}{H}{loa}
\providecommand{\algorithmname}{Algorithm}
\floatname{algorithm}{\protect\algorithmname}
\makeatother

\begin{document}

\title{Quantum Amplitude Estimation in the Presence of Noise}
\author{Eric G. Brown}
\thanks{These authors contributed equally to this work.}

\author{Oktay Goktas}

\author{W.K. Tham}
\thanks{These authors contributed equally to this work.}
\affiliation{Agnostiq Inc., 180 Dundas Street W, Suite 2500, Toronto, ON M5G 1Z8, Canada}\date{\today}

\begin{abstract}
Quantum Amplitude Estimation (QAE) -- a technique by which the amplitude of a given quantum state can be estimated with quadratically fewer queries than by standard sampling -- is a key sub-routine in several important quantum algorithms, including Grover search and Quantum Monte-Carlo methods. An obstacle to implementing QAE in near-term noisy intermediate-scale quantum (NISQ) devices has been the need to perform Quantum Phase Estimation (QPE) -- a costly procedure -- as a sub-routine. This impediment was lifted with various QPE-free methods of QAE, wherein Grover queries of varying depths / powers (often according to a ``schedule'') are followed immediately by measurements and classical post-processing techniques like maximum likelihood estimation (MLE). Existing analyses as to the optimality of various query schedules in these QPE-free QAE schemes have hitherto assumed noise-free systems. In this work, we analyse QPE-free QAE under common noise models that may afflict NISQ devices and report on the optimality of various query schedules in the noisy regime. We demonstrate that, given an accurate noise characterization of one's system, one must choose a schedule that balances the trade-off between the greater ideal performance achieved by higher-depth circuits, and the correspondingly greater accumulation of noise-induced error.

\end{abstract}
\maketitle

\section{Introduction}

Increasingly, the ability of classical computers to scale and keep pace with computational demands imposed by various modern economic, industrial, and creative pursuits is being strained. Quantum computing is an avenue through which a variety of computationally difficult problems may be solved more efficiently. Problems that may benefit from a quantum advantage span a broad spectrum ranging from well-known ones like integer factorization~\cite{Shor_1999} and search~\cite{Grover1997} to more modern techniques for molecular simulations~\cite{McArdle_2020, ryabinkin_2018} and machine learning~\cite{dunjko2017machine,Biamonte_2017,alanVQE,FarhiQAOA,FarhiQAOAsupremacy}. We are currently entering the so-called Noisy, Intermediate-Scale Quantum (NISQ) era of hardware \cite{Preskill_2018}, which describes the fact that current and near-term quantum computers are relatively small (few numbers of qubits) and are noisy and error-prone. The presence of noise means that practical circuit depths (number of quantum operations that can be sequentially performed) is severely constrained, which must be a principle factor of consideration when designing NISQ algorithms.

A central subroutine key to several quantum algorithms is Quantum Amplitude Estimation (QAE) \cite{brassard2000quantum}. In particular, QAE is used in Quantum Monte Carlo (QMC) methods, in which it gives a quadratic speed-up over classical Monte Carlo sampling. This can be used, for example, to perform simple integration \cite{Montanaro_2015}, and recently there has been considerable interest in using QMC to give advantage when pricing financial options and making risk analyses \cite{Rebentrost_2018, stamatopoulos2019option, Woerner_2019, egger2019credit}. Canonically, QAE uses Quantum Phase Estimation (QPE) as a subroutine \cite{brassard2000quantum}, necessitating the execution of the quantum Fourier transform and a large series of controlled unitaries that makes any NISQ implementation unlikely. Due to this limitation there have been several alternative approaches proposed that lets one perform QAE without the need to do QPE \cite{Suzuki_2020, grinko2019iterative, nakaji2020faster, Aaronson_2020}. These methods rely on performing and measuring a collection or series of circuits, the results of which are classically post-processed.

Even though these QPE-free QAE algorithms reduce the necessary circuit depth, it remains the case that all current methods that achieve quadratic improvements require circuits of rapidly increasing depth with increasing estimation accuracy, and thus these approaches are likely to be highly limited in the presence of system noise. Moreover, by the nature of the problem, one must characterize very precisely the noise present in one's system in order to account for it when making the amplitude estimation. Any bias resulting from a mischaracterization of the noise will produce a corresponding bias in one's estimate.

In this work we build on the MLE-QAE approach of~\cite{Suzuki_2020} to study the effects of some simple noise models on the performance of QAE. Among our main results, we will demonstrate that an ideal Heisenberg scaling of estimation error (i.e. a quadratic improvement over classical) either via the strategy proposed as optimal in~\cite{Suzuki_2020}, or standard QPE, is likely untenable on near-term hardware. On the other hand we show that there are strategies that are much more achievable at the expense of a less-than-quadratic quantum advantage in the asymptotic regime. Indeed, at modest circuit depths and similarly modest estimation precision -- often the regime of interest for near-term quantum devices -- certain strategies in the presence of noise are shown to outperform even asymptotically optimal strategies \emph{sans} noise.

Our work is organized as follows. We begin in Sect. \ref{sec:bg} with a background discussion on Quantum Amplitude Amplification and its use in quantum estimation, and give an overview of the MLE-QAE procedure specifically. We then move on to explain our methods in Sect.~\ref{sec:methods}; we begin with the definition of a toy QMC problem that we will be solving via QAE, and then delve into the topic of noise, its expected impact on amplitude estimation, and the specific noise models that we consider in this study. We will show here how to correct one's likelihood function to account for the noise models that we consider, and to construct a correspondingly modified Fisher Information and Cramer-Rao bound. In Sect.~\ref{sec:res} we show the results of our numerical studies, demonstrating how the ideal performance of QAE is impacted by different types and strengths of noise, in particular elucidating the QAE strategy to be used given a degree of noise. We then discuss these results in the context of current and near-term hardware. We finish with concluding remarks in Sect.~\ref{sec:con}. Three appendices are also included. Appendix.~\ref{A1} gives an introductory overview of maximum-likelihood methods, the Fisher Information, and the corresponding Cramer-Rao bound. Appendix.~\ref{B1} outlines the calculation by which to compute the noise-corrected Fisher Information, and Appendix.~\ref{A3} demonstrates how to compute states resulting from our noise models.

The reader should note that during the final stages of writing this manuscript there was a pre-print posted that shares several overlapping ideas and conclusions as those we posit here \cite{wang2020bayesian}, though through a different set of methods and perspective.

\section{Background} \label{sec:bg}

In preparation for considering noisy circuits, we begin in this section by giving the reader a review of QAE in the noise-free regime. A crucial step in QAE is the estimation of an eigenphase accrued upon multiple application of a so-called Grover operator, and as such
we also briefly survey various methods that have been developed for this phase estimation.

\subsection{Quantum Amplitude Amplification}

At heart, QAE is really a combination of Quantum Amplitude Amplification (QAA) -- a technique to amplify
the \emph{a priori} unknown amplitude of a ``marked'' state $\left|\psi_{0}\right>$ -- followed by
one of several possible techniques to estimate that amplified amplitude, hopefully with greater precision
than classically possible. The fundamentals of QAA, which we will now review, were originally described
by Brassard et al \cite{brassard2000quantum} in work that was in turn inspired by Grover's well-known search
algorithm \cite{grover1996fast}.

Let us start by supposing one is given a blackbox that prepares a quantum state: $\left|\psi\right>={\cal A}\left|0\right>$.
In general ${\cal A}$ can be any arbitrary unitary operator, but in the special case of Grover's search
algorithm it encodes the database over which the search is to be performed -- usually an equal superposition
over all computational basis states $\left|\psi\right>=\left|+\right>^{\otimes n}$ where $n$ is the
number of qubits being used in the encoding and $\left|\pm\right>=\frac{1}{\sqrt{2}}\left(\left|0\right>\pm\left|1\right>\right)$.
Now suppose that there is some ``good'' subspace spanned by set(s) of desirable states (say indices
corresponding to ``hits'' for a search algorithm). The projection of $\left|\psi\right>$ onto this
good subspace we will denote $\sqrt{\alpha}\left|\psi_{0}\right>$ so that in principle the encoded state can
be written as:
\begin{align}
\left|\psi\right> & =\sqrt{\alpha}\left|\psi_{0}\right>+\sqrt{1-\alpha}\left|\psi_{1}\right>\nonumber \\
 & =\cos\theta\left|\psi_{0}\right>+\sin\theta\left|\psi_{1}\right>\label{eq:BasicDecomposition}
\end{align}
where $\left|\psi_{1}\right>\propto\left|\psi\right>-\left|\psi_{0}\right>\left\langle \psi_{0}|\psi_{1}\right\rangle $.
Without loss of generality, we will always define $\left|\psi_{1}\right>$ such that $\alpha$ is real.

The goal of QAE ultimately is to estimate the amplitude $\alpha$.
Now, a simple means of accomplishing that goal is to just measure along the marked basis, say by implementing the
following projector-valued measure (PVM): $\{\left|\psi_{0}\right>\left<\psi_{0}\right|,\left|\psi_{1}\right>\left<\psi_{1}\right|\}$.
Doing so (over $N$ identical and independent -- or iid -- copies of $\left|\psi\right>$) yields an
uncertainty in estimates of $\alpha$ lower-bounded by the so-called standard quantum limit (SQL): $\sqrt{\mathbb{E}\left[(\alpha'-\alpha)^{2}\right]}\sim1/\sqrt{N}$
where $\alpha'$ is an estimate of $\alpha$.

A key observation in QAA, however, is that with access to repeated use of ${\cal A}$ (and its inverse
${\cal A}^{\dagger}$) along with queries to an oracle that can identify the marked state $\left|\psi_{0}\right>$,
one can modulate the encoded state $\left|\psi\right>$ such that $\theta$ is amplified, which we will
argue makes it easier to estimate. Consider the operator 
\begin{align}
\hat{\chi}_{\varphi} & =\mathbb{I}-2\left|\varphi\right>\left<\varphi\right|\label{eq:MirrorOp}
\end{align}
which imposes a sign flip on some selected state $\left|\varphi\right>$. Now suppose we are (a) provided
with some oracle $\hat{\chi}_{\psi_{0}}$ that acts on the special marked state $\left|\psi_{0}\right>$
and (b) with repeated use of ${\cal A}$ and ${\cal A}^{\dagger}$ we can implement ${\cal A}\hat{\chi}_{0}{\cal A}^{\dagger}=\mathbb{I}-2\left|\psi\right>\left<\psi\right|$
which acts on the encoded state $\left|\psi\right>$. Notice that both $\hat{\chi}_{\psi_{0}}$ and ${\cal A}\hat{\chi}_{0}{\cal A}^{\dagger}$
leave invariant the subspace spanned by $\left|\psi_{0}\right>$ and $\left|\psi_{1}\right>$. More explicitly,
defining the ``Grover" operator ${\cal Q}={\cal A}\hat{\chi}_{0}{\cal A}^{\dagger}\hat{\chi}_{\psi_{0}}$, one can show that
\begin{align}
\left|\psi^{(k)}\right> & ={\cal Q}^{k}\left|\psi\right>\nonumber \\
 & =\cos(2k+1)\theta\left|\psi_{0}\right>+\sin(2k+1)\theta\left|\psi_{1}\right>\label{eq:Qop}
\end{align}

Readers familiar with the metrological utility of various quantum states will recognize that $\left|\psi^{(k)}\right>$
is a more ``useful'' state compared to $\left|\psi\right>^{\otimes N}$, even though they ``cost''
the same number of queries ($N=2k+1$) to ${\cal A}$ (or ${\cal A}^{\dagger}$) to prepare. Formally,
we can quickly see this by noting that $\left|\psi^{(k)}\right>$ yields a quantum Fisher information~\cite{petzQfish} with respect to $\alpha$ of
\begin{align}
 & {\cal I}_{f}\left(\psi^{(k)},\alpha\right)\nonumber \\
= & 4\left\langle \partial_{\alpha}\psi^{(k)}\Big|\partial_{\alpha}\psi^{(k)}\right\rangle -4\left|\left\langle \partial_{\alpha}\psi^{(k)}\Big|\psi^{(k)}\right\rangle \right|^{2}\nonumber \\
= & 4\frac{\left(2k+1\right)^{2}}{\alpha(1-\alpha)}\label{eq:FishQ}
\end{align}
For a review of the Fisher information and the associated Cramer-Rao lower bound (CRB), the reader is referred to Appendix. \ref{A1}.
The CRB then immediately implies that unbiased estimates of $\alpha$ (if indeed it
is practical to construct) will exhibit errors \emph{lower bounded} by the Heisenberg scaling $\sqrt{\mathbb{E}\left[(\alpha'-\alpha)^{2}\right]}\sim1/\sqrt{{\cal I}_{f}}\propto1/N$.
If indeed this bound is tight, then there is a quadratic improvement over the classical $1/\sqrt{N}$
scaling in the naive case discussed above.

\subsection{Estimating the amplitude}

Historically, the \emph{estimation }part of QAE is accomplished by pairing QAA with quantum phase estimation
(QPE) \cite{Nielsen-Chuang}, leveraging the fact that the operator ${\cal Q}$ in
Eq.~\ref{eq:Qop} has eigenvalues of the form $e^{\pm2i\theta}$. Applying standard QPE with ${\cal Q}$
yields an efficient estimate of $2\theta$, which in turn allows us to deduce $\alpha$. Unfortunately,
relying on QPE has rather severe downsides in the context of near term quantum devices, as it relies heavily on the use of controlled versions of the operator ${\cal Q}$, as well as the use of the quantum Fourier
transform (QFT). Both of these require extensive use of two-qubit entangling gates which are physically challenging to implement with high fidelity, making even modest-sized problems untenable on near term devices.

A variety of schemes have been considered that eschew the use of QPE \cite{Suzuki_2020, grinko2019iterative, nakaji2020faster, Aaronson_2020}. These generally prescribe directly
measuring states that result from QAA, followed by post-processing. But while common techniques like
maximum-likelihood estimation (MLE) \cite{Suzuki_2020} or Bayesian inference approaches \cite{Wiebe_2016}
let us efficiently deduce $(2k+1)\theta$ for a \emph{particular} choice of $k$, one must be careful
to ensure that the ensemble of $k$'s produced via QAA in the first place is an adequate one. From our
preceding discussion around Eq.~\ref{eq:FishQ}, it might be tempting to prescribe preparing states
with the largest possible $k$ given the number of calls to ${\cal A}$ (or ${\cal A}^{\dagger}$ ) that
one is allowed, in order to maximize ${\cal I}_{f}$ and minimize the corresponding CRB. However, since
$\alpha$ is periodic in $\theta$, estimating $(2k+1)\theta$ (even when done with high precision) still
leaves one with an ambiguity as to \emph{which} $2\pi/(2k+1)$ interval $\theta$ actually belongs -- an ambiguity that does not afflict the state $\left|\psi\right>^{\otimes N}$.

In standard QPE, this is resolved by ensuring that an exponentially increasing sequence of multiples of $\theta$ is used at the input to the QFT stage, ensuring that the eventual estimate of $\theta$ is captured at different granularities.
In a similar spirit, Suzuki et al \cite{Suzuki_2020} recently calculated that if one were to perform
MLE upon measuring the output of QAA for an \emph{exponential} ``schedule'' of $k$'s (i.e. for $k=2^{s-1}$,
where $s=0,1,2,3,...$ and $k_0=0$), one asymptotically recovers a Heisenberg scaling in estimates
of $\theta$, without ambiguities stemming from the $2\pi$ periodicity of $\alpha$. By contrast, a
similar calculation for a \emph{linear} schedule (i.e. for $k=0,1,2,...$ with equal weights), yields an error scaling that is intermediate between classical and Heisenberg: $\sqrt{\mathbb{E}\left[(\alpha'-\alpha)^{2}\right]}\sim1/N^{(3/4)}$.
More generally, a polynomial schedule of degree $d$ (i.e. $k=s^d$, where $s=0,1,2,3...$) yields the asymptotic (in $N$) scaling $\sqrt{\mathbb{E}\left[(\alpha'-\alpha)^{2}\right]}\sim1/N^{(2d+1)/(2d+2))}$

Note that in quantum order- and factor-finding algorithms, the operator ${\cal Q}^{2^{s}}$ for larger $s$ can be made asymptotically efficient via modular exponentiation \cite{Nielsen-Chuang}. This trick does not generally apply outside of the standard usage of QPE, however, for instance with the exponential schedule of $\mathcal Q$ operators considered by Suzuki et al.

%Such a procedure is itself resource-intensive in the context of near-term devices so a schedule such as Yamomoto et al's has query complexity that is exactly ${\cal O}(k)$ and is therefore exponential in the schedule index $s$.

In this work, we consider the effect of noise on the feasibility of various MLE schedules. In particular,
noting that because noise compounds quickly with circuit depth, an exponential schedule that may be asymptotically
optimal in the ideal case quickly becomes suboptimal in the presence of noise. We also proceed to explore
alternate schedules and their optimality in various noise regimes.

\subsection{QAE via MLE}  \label{MLE-QAE}
In practical usage, it is often convenient (whenever the encoding allows it) to mark the ``good'' subspace onto which $\ket{\psi}$ is being projected with a ``readout'' ancilla qubit so that Eq.~\ref{eq:BasicDecomposition} now reads:
\begin{align} \label{BasicDecompositionReadout}
    \ket{\psi}=\cos\theta\ket{\psi_{0}}\ket{0}+\sin\theta\ket{\psi_{1}}\ket{1},
\end{align}
where the amplitude is again $\alpha = \cos^2 \theta$. Doing so allows one to easily realize the operator $\hat{\chi}_{\psi_{0}}$ as well as the PVM $\{\left|\psi_{0}\right>\left<\psi_{0}\right|,\left|\psi_{1}\right>\left<\psi_{1}\right|\}$.

Now suppose we performed the PVM $\{\ket{0}\bra{0},\ket{1}\bra{1}\}$ on states prepared in accordance to some schedule $\{\mathcal{Q}^{k_1}\ket{\psi},\mathcal{Q}^{k_2}\ket{\psi},\mathcal{Q}^{k_3}\ket{\psi},...\}$. Assume that for each $k_s$ the measurement is performed over $N_s$ iid shots, and let $p_s$ be the fraction of ``good" measurements, which we know asymptotically has the form $p_{s} \xrightarrow {N_s \rightarrow \infty}\left|\left\langle 0\big|{\cal Q}^{k_{s}}\big|\psi\right\rangle \right|^{2}$.
Given an observation $p_s$, the likelihood function is a function of $\theta$ and given by 
\begin{align}
     & \mathcal{L}_{s}(\theta|p_{s})\nonumber \\
    = & \left[\cos^{2}(2k_{s}+1)\theta\right]^{N_{s}p_{s}}\left[\sin^{2}(2k_{s}+1)\theta\right]^{N_{s}(1-p_{s})}\label{eq:Lk}
\end{align}
For a brief review of the likelihood function, the Fisher Information, and the Cramer-Rao bound, the reader is referred to the Appendix. \ref{A1}. Overall, the likelihood function resulting from all experiments in a schedule (and the likelihood to be maximized in order to estimation $\alpha$) is:
\begin{align} \label{L}
    \mathcal{L}(\theta|\{p\})=\prod_{s=0}^{s_{\text{max}}}\mathcal{L}_{s}(\theta|p_{s}),
\end{align}

In the remainder of this paper, we use this simple MLE framework to explore alternate schedules $\{k_s\}$ in the context of noisy circuits; in addition to linear and exponential schedules we also consider polynomial schedules, for example the quadratic schedule $k_s=s^2$. As is standard in the QAE literature, for an apples-to-apples comparison between schedules we normalize the performance of each by the total number of calls to the encoding operator $\mathcal{A}$ or $\mathcal{A^\dagger}$, such that:
\begin{align}
    \sum_{s=0}^{s_{\text{max}}}2k_s+1
\end{align}
is constant between two schedules when comparing their performance.

\section{Methods}     \label{sec:methods}
In this section, we will begin by describing a toy problem that we use as a prototypical application of QAE in Section~\ref{sec:toy}.
We then briefly allude in \ref{questions} to the question of noise and its expected impact on the problem, before discussing our studied noise models and solution methods in \ref{sec:noises}.

\subsection{Toy Problem for Quantum Monte Carlo}  \label{sec:toy}

Here and throughout the remainder of this text we will be using QAE to solve a simple toy problem, namely, computing the expectation value $\mathbb{E}(\cos^2 x)$ over an interval $[0, b_\text{max})$. That is, solving the integral
\begin{align} \label{I}
    S = \frac{1}{b_\text{max}} \int_0^{b_\text{max}} \cos^2 x \, dx = \mathbb{E}(\cos^2 x),
\end{align}
where the expectation value is taken with respect to the uniform measure over the interval $[0, b_\text{max})$.
Unless stated otherwise, the default value is chosen to be $b_\text{max} = \pi/5$. 
% \et{Cute example... though somewhat arbitrary so I guess $b_{max} = \pi/5$ is as good as anything else. I wonder if we should dwell ont he noise-free performance as much as we do at all though, given that's already been discussed elsewhere... I mean OK, not with respect to this exact example, but still...}

% \eb{Regarding the problem and value of $b_\text{max}$, recall that I originally started with some code in the qiskit community tutorials. They use this problem and that value of $b_\text{max}$ in the notebook, which is why we may want to change $b_\text{max}$. I think it's fine to use the problem, as it's basically the simplest problem possible to be solved with QMC, and so makes a good standard for benchmarking.}

% \eb{Regarding the noisless results: I agree. My idea was to highlight the polynomial schedules and show that, even without noise, a polynomial schedule will often work equally well or even better than an exponential schedule, depending on your desired error. How important of a result do we think that is to highlight? I can certainly reduce this though, and maybe put in a simple scaling analysis for polynomial schedules instead, as done with linear and exponential.}
In classical Monte-Carlo integration, $S$ can be calculated by sampling uniformly over $x\in[0,b_{max})$, computing $\cos^2(x)$ for each sample and then averaging.
An analogous quantum approach uses the states of $m$ qubits to represent a discretization of the interval into $2^m$ bins. Uniformly sampling over the interval then corresponds to preparing the state $\ket{x}=\ket{+}^{\otimes m}$, and the encoding of the amplitude $\mathbb{E}(\cos^2x)$ is then performed by applying controlled rotations $\bigotimes_{j=1}^{m}\hat{R}_{y}^{x_{j}}(b_{max}/2^{j})$ targeting an ancilliary readout qubit (see \cite{stamatopoulos2019option} for an excellent exposition of the procedure).
Here, $\hat{R}_{y}(\theta)=\exp(i\hat{\sigma}_{y}\theta)$, and $x_j$ is the (binary) value of the $j$-th qubit of the $\ket{x}$ register.
In other words, in the language of QAA, the encoding operator prepares:
\begin{align}
 & {\cal A}\ket{0}\ket{0} \nonumber \\
= & \sum_{x=0}^{2^{m}-1}\ket{x}\left[\cos\left(\frac{b_{\text{max}}x}{2^{m}}\right)\ket{0}+\sin\left(\frac{b_{\text{max}}x}{2^{m}}\right)\ket{1}\right],
\end{align}
such that the expectation value of $\ket{0}\bra{0}$ on the readout qubit gives us the desired expectation value of $\mathbb{E}(\cos^2 x)$, evaluated on an $m$-bit discretization of the interval $[0,b_{max})$.
Having defined $\cal{A}$, one can now call upon QAA and QAE as discussed in Section~\ref{sec:bg} to improve one's estimate of the expectation value.
%  Numerically one can estimate the value of the true average by sampling uniformly over $x$ within the interval and computing the numerical average of $\sin^2 x$ over the samples, i.e. through a Monte Carlo procedure. We will solve it in the same manner but by utilizing QAE to perform Quantum Monte Carlo. This is done by discretizing the region $[0, b_\max]$ into $2^n$ regions and represent these regions as basis states of an $n$-qubit register. Note that this means we are actually estimating a Riemann summation approximation of Eq. (\ref{I}). Uniformly sampling over $x$ then corresponds to the uniform superposition over all regions, created by applying Hadamard gates to all qubits. The subsequent Quantum Monte Carlo procedure is outlined in Appendix. \ref{A2}.

\subsubsection{Noise-Free Schedule Performance}

\begin{figure}[t]
	\centering
    \includegraphics[width=0.50\textwidth]{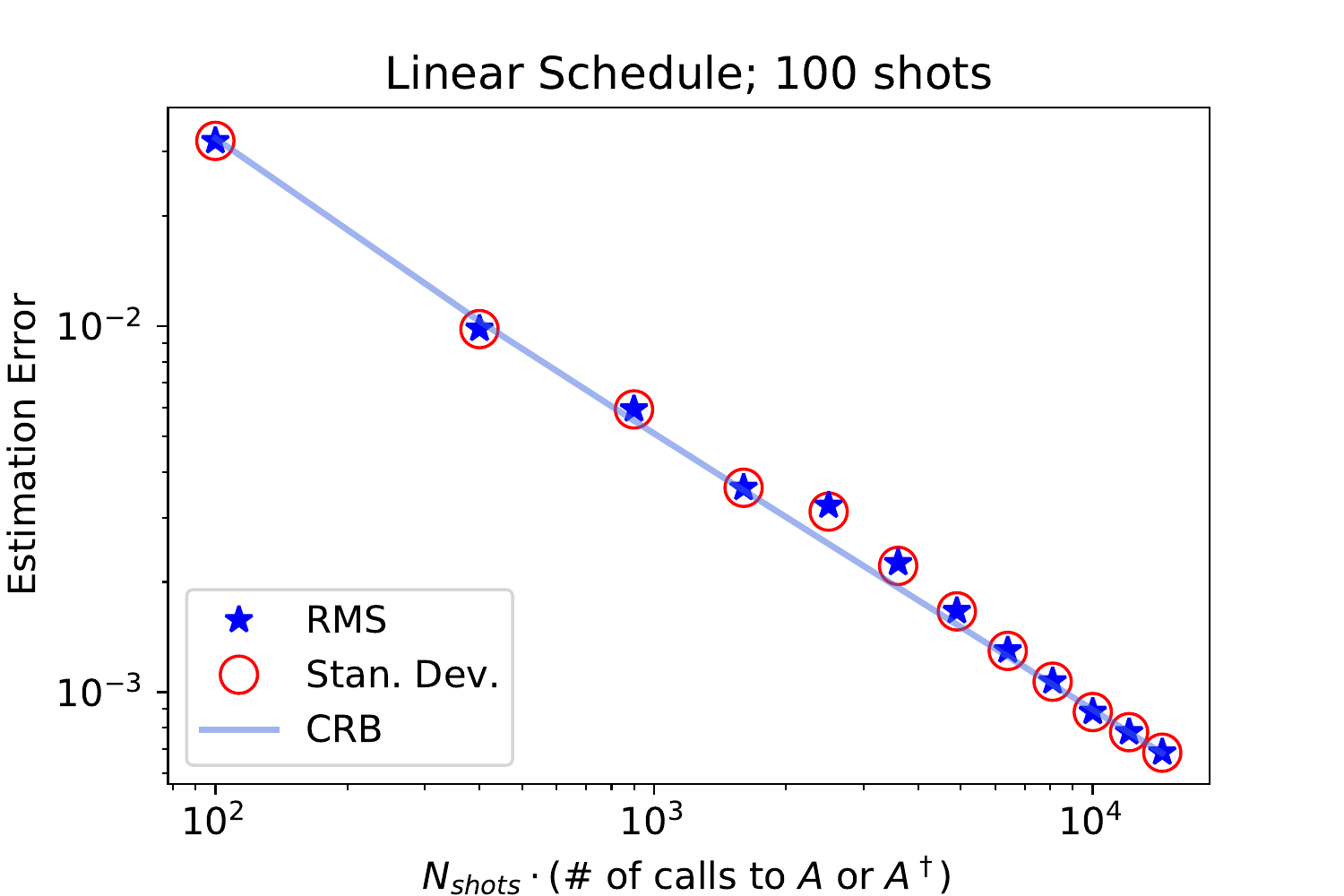}
	\caption{A linear schedule with $N_\text{shots}=100$ shots per experiment. We plot the RMS estimation error as well as the standard deviation and CR bound. We see that RMS error agrees very well with the standard deviation, and that they both follow the CR bound tightly, indicating that $100$ shots is sufficient to reduce the bias to a negligible value and put us in the asymptotic regime.}
	\label{bias2}
\end{figure}

In our simulations we chose to perform $N_\text{shots}=100$ shots per experiment in a given schedule. To confirm that this is sufficient to put us into an asymptotic regime (meaning that the RMS error of estimation numerically conforms with the predictions of the Cramer-Rao bound), we perform a QAE simulation using a linear schedule and plot in Fig. \ref{bias2} the RMS error, CR bound, as well as standard deviation of error for our estimation. The linear schedule is defined as $k_s =s$. The fact that the standard deviation and RMS error overlap indicates that the bias in our estimation error is negligible, and thus that the CR bound can be used as an accurate indicator of performance.

\begin{figure}[t]
	\centering
    \includegraphics[width=0.5\textwidth]{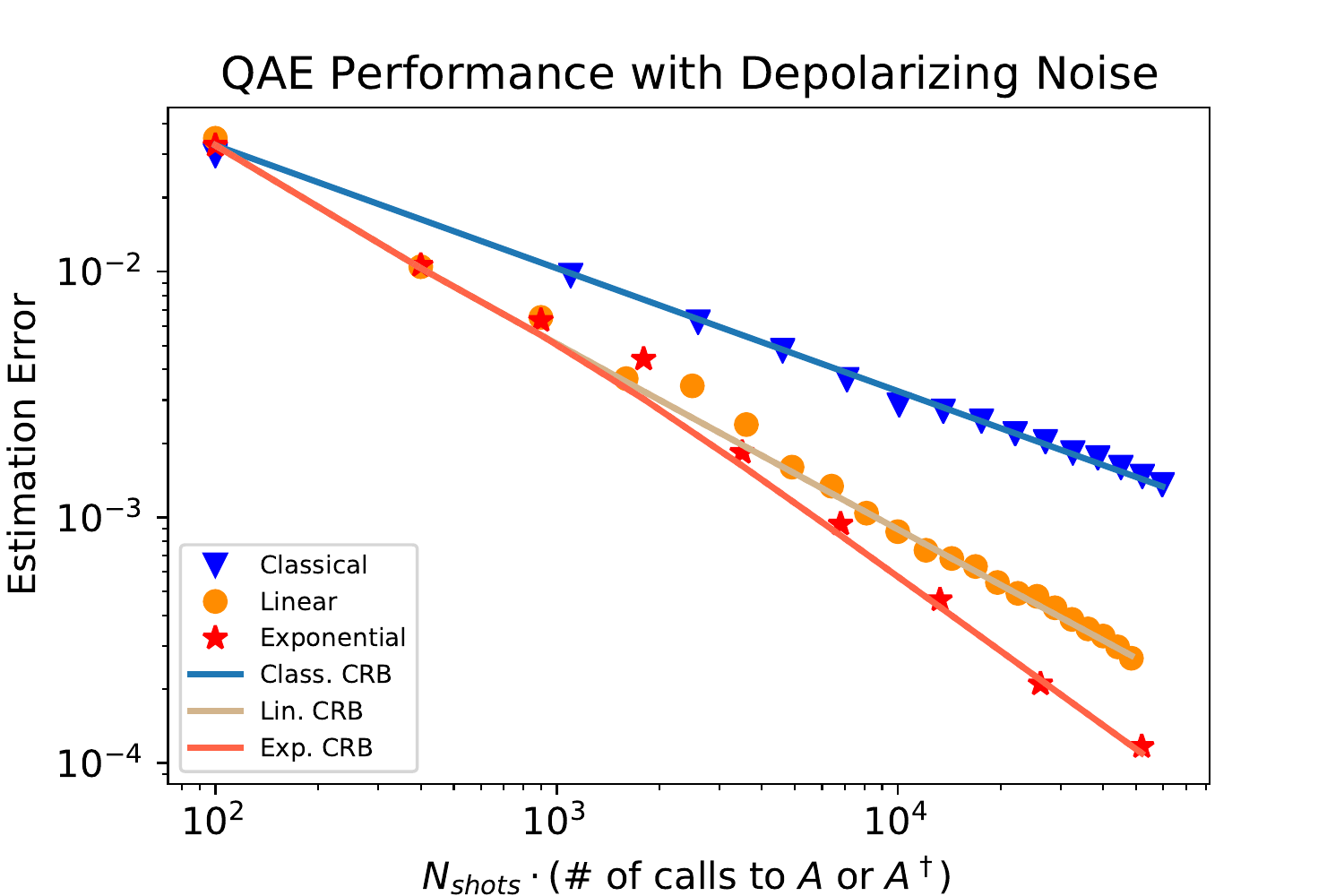}
	\caption{The RMS-error performance of classical, linear, and exponential schedules with $N_\text{shots}=100$ shots in the absence of noise, along with their corresponding CR bounds.}
	\label{Clean}
\end{figure}

As a benchmark for the noisy case, in Fig. \ref{Clean} we compare the RMS-error performance of this linear schedule with that of classical sampling and the (Heisenberg-bound achieving) exponential schedule, $k_s = 2^{s-1}$ with $k_0=0$. We observe the expected scalings as demonstrated in \cite{Suzuki_2020}. Note that, as they are defined, the first three experiments of the linear and exponential schedules are equivalent, with number of Grover operators $k_s = 0, 1, 2$.

%\begin{figure}[t]
%	\centering
%    \includegraphics[width=0.45\textwidth]{Noisless1.png}
%	\caption{Classical, linear, and exponential.}
%	\label{Noisless1}
%\end{figure}
%\begin{figure}[t]
%	\centering
%    \includegraphics[width=0.45\textwidth]{Noisless.png}
%	\caption{Includes quadratic}
%	\label{Noisless}
%\end{figure}

\subsection{Questions of Noise and Depth} \label{questions}

We have seen that an exponential Grover schedule achieves the optimal Heisenberg scaling. However, such a scheme also means that a large fraction (half here) of the total number of applications of $\mathcal A$ and $\mathcal A^\dagger$ will take place in series in a single, exponentially increasingly large-depth circuit, which in practice will quickly fail on real near-term devices. This motivates the exploration of the effects of noise on the MLE-QAE procedure, and in particular elucidate the optimal schedule in practice, given a noise model and a maximum effective circuit depth.

It is worth noting that the other QAE procedures which achieve the Heisenberg bound \cite{brassard2000quantum, grinko2019iterative, nakaji2020faster, Aaronson_2020}, including QAE with phase estimation, also all involve implementing a series of Grover circuits with exponentially increasing depth, and thus will suffer the same scaling with noise as does the exponential schedule in ML-QAE. The implication of noise thus equally applies to all other known QAE procedures, and we can expect qualitatively similar impacts on their performance due to noise.

\subsection{Noise Models} \label{sec:noises}

The presence of noise in quantum computers is one of the fundamental limiting factors on the extent of quantum algorithms that can realistically be implemented, and thus it behoves us to understand the ways in which Quantum Amplitude Estimation is limited by noise. In doing so we will here restrict ourselves to noise models in which only the ancillary measurement qubit experiences noise. Making this choice simplifies the computations of the likelihood functions to be optimized, and of the Fisher Information and consequent Cramer-Rao bounds.

This assumption of ancilla-only noise may actually not be a poor approximation to reality when using a larger number of qubits $n$, as likely to be the case in any real application of QAE. This is because of the form of the Grover operator $\mathcal{Q}$, which contains many two-qubit gates between the ancilla and the other $n$ qubits. The number of these two qubit gates in our problem (interactions experienced by the ancilla over the course of a Grover operation) can be shown to scale with $n$, and thus the strength of noise impacting the ancilla will generally grow exponentially with $n$. Contrast this with a given qubit in the $n$-register, which experiences a constant number of interactions between itself and the ancilla (not increasing with $n$). For large $n$ one should therefore expect the noise on these qubits to be negligible compared to the ancilla.

\begin{figure}[t]
	\centering
    \includegraphics[width=0.45\textwidth]{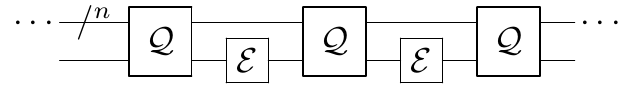}
	\caption{The form of noise models that we consider on series of Grover operations. We will consider the cases in which the noisy channel $\mathcal{E}$ corresponds to depolarizing, dephasing, and amplitude damping channels.}
	\label{circuit}
\end{figure}

We further simplify our noise models to consist of single noisy channels applied to the ancilla after every Grover operation. That is, given a single-qubit channel $\mathcal{E}$, we simulate circuits consisting of Grover series of the form in Fig. \ref{circuit}. Below we will consider channels $\mathcal E$ corresponding to depolarizing, dephasing, and amplitude damping noise.

%\eb{@Edwin do you think it's worth it to mention something like.. "In future generations of quantum hardware it may be that the state preparation / oracle call may be off-shored to a designated quantum device optimized for this task, and that sources of error are more likely to occur from the interaction between this device and the ancilla qubit/device..."?}

\subsubsection{Depolarizing Noise} \label{depolSect}

The first and simplest noise model that we consider is that of depolarizing noise. The depolarizing channel, applied to a readout state $\rho_r$, produces
\begin{align}
    \mathcal{E}(\rho_r) = (1-\gamma)\rho_r + \gamma\frac{\mathbb I_2}{2}, \label{DepolAction}
\end{align}
where $\mathbb I_2$ is the $2\times2$ identity and $\gamma$ is the probability that the state $\rho_r$ is replaced with the maximally mixed state.
%That is, with probability $\gamma$ the state is replaced by the fully mixed state.
%This model represents a good starting point and benchmark for studying noise, as for a given error rate it represents the informationally worst-case scenario (if an error occurs then all state information is lost).
The depolarising channel is a good starting noise model, in part because it is convenient to analyze.
In fact, as we will see in Sect. \ref{sec:CRB}, we rely on decompositions of other channels into a mixture between a depolarising one and a unitary operator in order to ease our calculation of QAE performance.

Given the action of depolarising noise, determining the effective noise resulting from chaining $k$ noisy Grover operations (as in Fig. \ref{circuit}) is particularly easy since the fully mixed state on the RHS of Eq.~\ref{DepolAction} remains invariant under additional unitary operations.
Consecutive applications of noisy (depolarising) Grover operators therefore behaves simply like a sequence of clean/noiseless Grover operators followed by a single stronger (greater $\gamma$) depolarising channel.
Strictly speaking, the latter statement is only true because (a) the readout qubit is always a target rail in interactions (mediated by two-qubit gates) with the $n$-qubit encoding rail and (b) we only measure the readout qubit and trace away the $n$-qubit encoding register.
Otherwise, noise in the readout qubit alone can manifest in the full $n+1$-qubit Hilbert space as non-trivial correlations that may be difficult to compute (and indeed can sometimes be useful)~\cite{KnillLaflammeDQC1}.

Now suppose we were to chain $k_s$ Grover operators (interspersed with an equal number of depolarising channels), the \emph{effective} depolarising strength is simply
%Furthermore, in a Grover operation the interactions between the ancilla and $n$-register consist purely of controlled operations in which the ancilla is the target, and as such the state of the ancilla has no back-action on the $n$-register, meaning that it will remain in the maximally mixed state. Thus, $k$ depolarizing Grover operations reduces to $k$ noiseless Grover operations followed by a single depolarizing channel in which the probability of no error is compounded to $(1-\gamma)^k$, giving a probability of error
\begin{align}
    \gamma_{k_s} = 1 - (1-\gamma)^{k_s}.
\end{align}
%Therefore the amplitude that one will measure from the ancilla after $k$ such operations, $\widetilde{\alpha}_k$, is given by 
The resulting probability ($\tilde{\alpha}_s$) with which a subsequent measurement of the readout register $\rho_r$ in the computational basis yields $0$ is:
\begin{align} 
    %\widetilde{\alpha}_k = (1-\gamma_k) \alpha_k + \frac{\gamma_k}{2},
    \widetilde{\alpha}_s = (1-\gamma_{k_s}) \alpha_{s} + \frac{\gamma_{k_s}}{2}
\end{align}
%where $\alpha_k = \sin^2 ((2k+1)\theta)$ is the corresponding noiseless amplitude. 
where $\alpha_{s} = \cos^2 ((2k_s+1)\theta)$ is the corresponding probability in the \emph{absence} of noise.
This modified measurement outcome allows us write down a new likelihood function for the $s$-th (noisy) experiment in the schedule as 
%Consider $m_k$ successive applications of the noisy Grover operator, corresponding to $D_k \equiv 2m_k$ oracle calls (here we ignore the $+1$ corresponding to the preparation of $\ket{\psi}$, assuming it to be noiseless). Depolarizing noise then reduces to a single depolarizing noise channel of the same form but with compounded error-free probability $p_k = p^{2m_k}$, and with probability of error $1-p_k$. Since the probabilities $p_k$ exponentially decreases with oracle circuit depth $D_k$, let us express it as
%\begin{align}  \label{pk}
%    p_k = \exp(-D_k/D_\text{err}), 
%\end{align}
%where $D_\text{err}$ is a free parameter describing the characteristic depth limit of a given device or noisy simulator. Below we will often refer to the depolarizing \emph{noise parameter} as the value of $1/D_\text{err}$
%With the modified amplitude, we can construct new likelihood functions accounting for this noise, to be used in the MLE-QAE scheme as described in Sect. \ref{MLE-QAE}. Given a schedule $\{k_s\}$, with corresponding noisy amplitudes $\widetilde{\alpha}_s$, the likelihood for experiment $s$ is
\begin{align} \label{noisyL}
    \widetilde{\mathcal{L}}_s = \widetilde{\alpha}_s^{N_s p_s}(1-\widetilde{\alpha}_s)^{N_s(1-p_s)}.
\end{align}

From this modified likelihood the MLE-QAE procedure as discussed in Sect. \ref{MLE-QAE} follows. Furthermore the corresponding Fisher information is also readily computable. For details on this calculation the reader is referred to Appendix.~\ref{B1}. In summary,
using additivity of the Fisher information for multiple independent measurement outcomes, the overall Fisher information from $N_s$ iid. shots for each entry $s$ in our schedule is computed to be
\begin{align}
     & \mathcal{\tilde{I}}_{f}(\alpha)\nonumber \\
    = & \frac{1}{\alpha(1-\alpha)}\sum_{s}(1-\gamma)^{2k_{s}}\,\frac{\alpha_{s}(1-\alpha_{s})}{\widetilde{\alpha}_{s}(1-\widetilde{\alpha}_{s})}\cdot N_{s}(2k_{s}+1)^{2},\label{noisyF}
\end{align}
from which the corresponding Cramer-Rao bound follows.

\subsubsection{Other Noise Models} \label{dephasingSect}

We will also consider the case in which $\mathcal E$ is the amplitude damping channel, the action of which on the readout qubit can be written in the Kraus operator formalism as $\mathcal{E}(\rho)=E_{0}\rho E_{0}^{\dagger}+E_{1}\rho E_{1}^{\dagger}$ where the Kraus operators $E_0$ and $E_1$ are defined as
\begin{align}
    E_0 = 
    \begin{pmatrix}
        1 & 0 \\
        0 & \sqrt{1-\gamma}
    \end{pmatrix}, \;\;\;
    E_1 =
    \begin{pmatrix}
        0 & \sqrt{\gamma} \\
        0 & 0
    \end{pmatrix}.
\end{align}
With probability $\gamma$, an input state is replaced with the ground state $\ket{0}$.

Finally, we consider dephasing noise, which contracts the Bloch sphere in every direction except along the computational basis. The action of the channel is described by the Kraus operators
\begin{align}\label{eq:defDephasing}
    E_0 = 
    \begin{pmatrix}
        1 & 0 \\
        0 & \sqrt{1-\gamma}
    \end{pmatrix}, \;\;\;
    E_1 =
    \begin{pmatrix}
        0 & 0 \\
        0 & \sqrt{\gamma}
    \end{pmatrix}.
\end{align}
That is, with probability $\gamma$ the ancilla undergoes complete dephasing.

Like depolarising noise discussed in the previous section, a damping channel does \emph{not} admit a decoherence free basis -- every possible preparation eigenbasis for $\rho_r$ undergoes equal contraction.
Unlike depolarising noise, however, the replacement ``information-less'' state in this case is the ground state, which in general is \emph{not} invariant to subsequent unitary operations.

On the other hand, we included dephasing noise in our analysis because it \emph{does} have a ``clean'' basis -- the channel leaves polar states ($\ket{0}$ and $\ket{1}$) untouched.
Therefore, as the readout qubit precesses between $\ket{0}$ or $\ket{1}$ and $\ket{\pm}$ with additional applications of $\mathcal Q$, it oscillates between being maximally degraded by dephasing and being unaffected by it.
One consequence is that for a given circuit depth (i.e. a given number of $\mathcal{Q}$ calls) the probability that any encoded information will not have dissipated (which we will make concrete and denote by $p_{\text{eff}}$ in the next section) is generally greater than in the fully-destructive case of depolarizing noise.

\subsubsection{Efficiently Computing the CRB given noise} \label{sec:CRB}

% Restricting ourselves to noise that afflicts only the readout qubit significantly simplifies our analysis. Under the single-qubit noise channels that we consider, the readout qubit stays within the\emph{ }$d=2$ subspace that it would have occupied in the absence of noise.
As we have already seen, analyzing the effects of chained noisy channels (as illustrated in Fig.~\ref{circuit}) is easy with depolarising noise.
Such is not the case with the other noise models under consideration.
Nevertheless, we can still write the outcome of our noise channels as a convex decomposition into a pure state and the maximally mixed state (on the readout qubit), even if it is not a ``natural'' decomposition for the channel in question. We denote this
\begin{align}
\mathcal{E}\rho_{r}\mathcal{E}^{\dagger} & =pU_{\psi_{r}}^{(\mathcal{E})}\left|\psi_{r}\right>\left<\psi_{r}\right|U_{\psi_{r}}^{(\mathcal{E})\dagger}+(1-p)\frac{\mathbb{I}}{2},
\end{align}
where now $1-p$ is interpreted as the probability of error. Here, $U_{\psi_{r}}^{(\mathcal{E})}$ is an \emph{effective} single-qubit unitary operator that depends on the channel $\mathcal{E}$ and its action on the (pure) readout state $\left|\psi_{r}\right>$. As an example, if $\left|\psi_{r}\right>=\left|+\right>$ and $\mathcal{E}$ is the dephasing channel, then $U_{\psi_{r}}^{(\mathcal{E})}$ is simply the identity operator.

A simple (if a little tedious) decomposition like this allows us to easily chain channels as shown in Fig.~\ref{circuit}, subject to the caveats mentioned in Sect.~\ref{depolSect}. The outcome of many applications of the $\mathcal{Q}$ operator followed by noise channel(s) can be compactly described as:
\begin{align}
\rho_{\text{eff}}= & \left(\prod_{s=1}^{m}\mathcal{\mathcal{Q}\mathcal{E}}\right)\rho_{r}\left(\prod_{s=1}^{m}\mathcal{\mathcal{Q}\mathcal{E}}\right)^{\dagger} \nonumber \\
= & \left(\prod_{s=1}^{m}p_{s}\right)\left[U^{(\mathcal{E},m)}\ket{\psi_{r}}\bra{\psi_{r}}U^{(\mathcal{E},m)}{}^{\dagger}\right] \nonumber \\
 & +\left(1-\prod_{s=1}^{m}p_{s}\right)\frac{\mathbb{I}}{2} \nonumber \\
= & p_{\text{eff}}\left|\psi_{\text{eff}}\right>\left<\psi_{\text{eff}}\right|+(1-p_{\text{eff}})\frac{\mathbb{I}}{2}
\end{align}
where in the last line we have used the fact that the identity matrix $\mathbb{I}$ is invariant under conjugation by unitary operators, a point we have belaboured in Sec.~\ref{depolSect}. Here, $U^{(\mathcal{E},m)}=\prod_{s=1}^{m}U_{s}^{(\mathcal{E})}Q$ is some \emph{effective} unitary given $m$ appplications of $\mathcal{Q}$ and $\mathcal{E}$.

Directly computing the sequence of \emph{effective }unitary operations $U_{s}$ is, in general, non-trivial (with the exception of the depolarising). However, for any given starting readout state $\left|\psi_{r}\right>$, it is easy to compute the trajectory of the output state $\rho_{\text{eff}}$ after $m$ applications of $\mathcal{Q}$ (and corresponding noise channels). The pure state portion, which we'll denote
\begin{align}
\left|\psi_{\text{eff}}\right> & =U^{(\mathcal{E},m)}\left|\psi_{r}\right> \nonumber \\
 & =\cos\varphi\left|0\right>+\sin\varphi\left|1\right>,
\end{align}
and the probability $p_{\text{eff}}$ that this pure state survives the noise channels allows us to compute the desired Fisher information and CRB directly.

Following the same procedure as for a generic depolarising channel (as outlined in Appendix. \ref{B1}), the Fisher information for $\alpha$ stemming from measuring the noisy state $\rho_{\text{eff}}$ in the computational basis is:
\begin{align}
\mathcal{I}_{f}= & 4\left(\frac{\partial\varphi}{\partial\theta}\right)^{2}\frac{p_{\text{eff}}^{2}\sin^{2}\left(\varphi\right)\cos^{2}\left(\varphi\right)}{\sin^{2}2\theta}\times\nonumber \\
 & \Bigg(\frac{1}{p_{\text{eff}}\cos^{2}\left(\varphi\right)+0.5(1-p_{\text{eff}})}\nonumber \\
 & +\frac{1}{p_{\text{eff}}\sin^{2}\left(\varphi\right)+0.5(1-p_{\text{eff}})}\Bigg).\label{eq:FisherNoisy}
\end{align}
Once again, because $\mathcal{I}_f$ is additive, we must sum Eq.~\ref{eq:FisherNoisy} over all schedule entries (which determine $\varphi$) and number of shots per entry.
Note that here $\partial{\varphi}/\partial{\theta}$ is non-trivial to compute in general and depends on the noise channels $\mathcal E$. As seen in Eq.~\ref{noisyF}, for depolarizing noise (as well as the noise-free case) this simply reduces to $(2k_s+1)$.

\begin{figure}
\begin{centering}
\includegraphics[width=8cm]{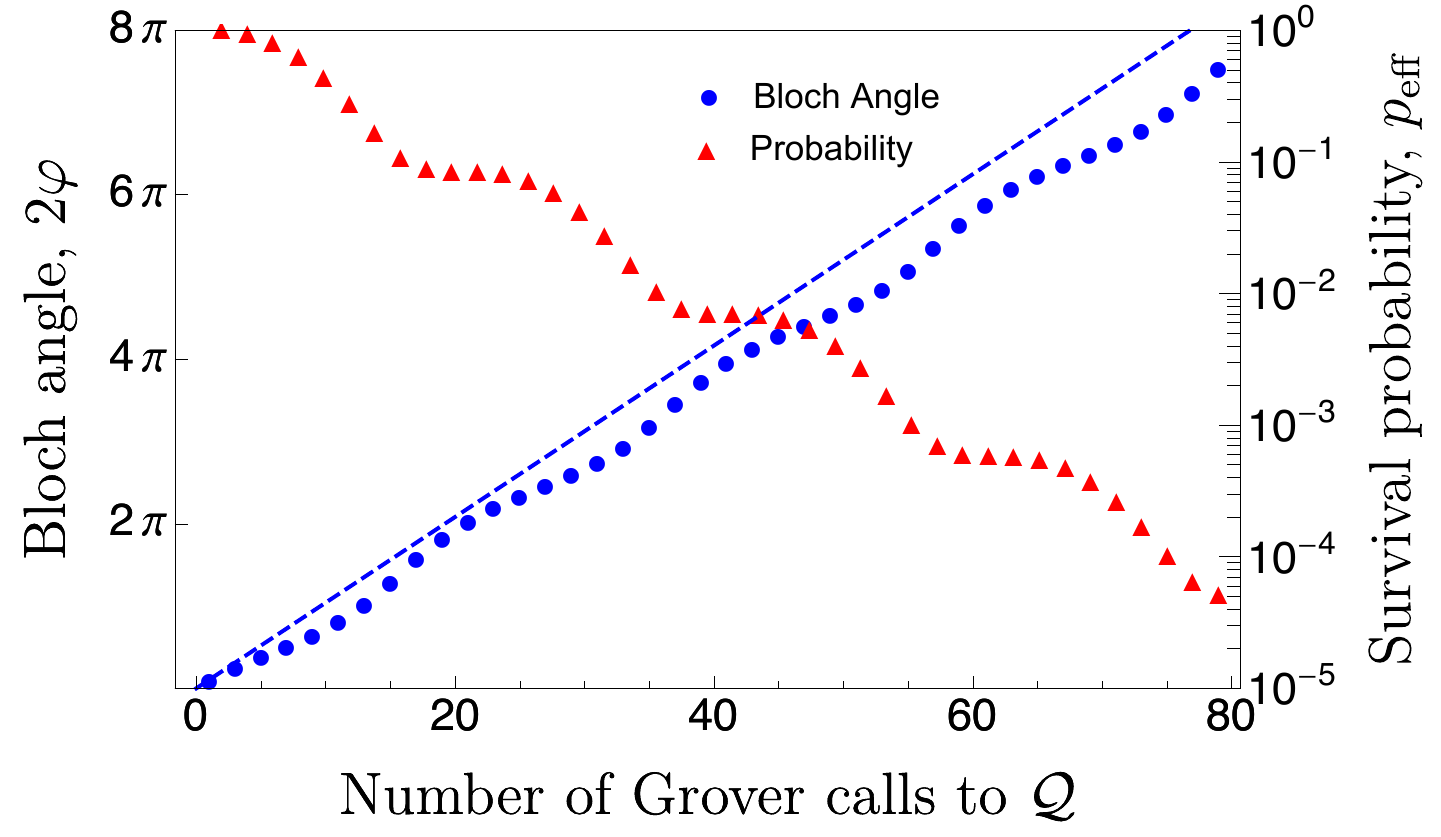}
\par\end{centering}
\caption{Plot of trajectory of pure state $\psi_{\text{eff}}$ and the probability $p_{\text{eff}}$ for that pure state to survive all noisy channels up to the stated number of applications of Grover operator $\mathcal{Q}$. The dashed line represents the analogous Bloch angle for the case of depolarizing noise. \label{fig:DampingTrajectory} }
\end{figure}

For a visual illustration, Fig.~\ref{fig:DampingTrajectory} shows an example of the trajectory of $\varphi$ and $p_{\text{eff}}$ under an amplitude damping channel with $\gamma=0.2$.
The salient feature to notice is that $\varphi$ deviates from the noiseless (and indeed the depolarising noise) case -- which is $(2k_s+1)\theta$ and is represented by the dashed line -- in a non-trivial way.
Damping noise here increases $\varphi$ (and decreases the purity $p_\text{eff}$) more quickly when $\varphi$ is such that $\ket{\psi_\text{eff}}$ is closer to the excited state $\ket{1}$ (i.e. $\pi/2\leq 2\varphi (\text{mod }2\pi)\leq 3\pi/2$).
In Appendix~\ref{A3}, we describe in further detail how $\varphi$ might be efficiently computed for any given $\theta$, $k_s$, and noisy channel $\mathcal E$.

\subsubsection{Aliasing and alternate schedule strategies} \label{aliasing}

A particularly pernicious effect of the noise models that we considered is the fact that certain numbers of calls to $\mathcal{Q}$ result in a state $\left|\psi_{\text{eff}}\right>$ with Bloch angles that are almost polar (i.e. $\varphi\approx n\pi$ for $n\in\mathbb{Z}$). In the absence of noise (i.e. setting $p_{\text{eff}}=1$ in Eq.~\ref{eq:FisherNoisy}) the Fisher information is a constant (with respect to $\varphi$) that depends only on the slope $\partial\varphi/\partial\theta$. In turn, as we have seen, this slope increases with the number of applications of $\mathcal{Q}$.

In the noisy case however (i.e. $p_{\text{eff}}<1$), the Fisher information becomes crucially dependent on $\varphi$; if a particular sequence of $\mathcal{Q}$ operators and noise channels happens to produce $\left|\psi_{\text{eff}}\right>\approx\left|0\right>$ or $\left|\psi_{\text{eff}}\right>\approx\left|1\right>$, the Fisher information plummets significantly, and vanishing entirely when $\varphi$ is exactly an integer multiple of $\pi$. This is precisely the effect behind the ``stair-stepped'' appearance of the Cramer-Rao lower bound (CRLB) in Figures~\ref{manyDepols}, \ref{manyDephase}, and \ref{manyDamping} in the presence of noise -- additional applications of $\mathcal{Q}$ is not guaranteed to improve the precision of our estimate of $\theta$ if, in tandem with the effects of intervening noise channels, it happens to yield a $\left|\psi_{\text{eff}}\right>$ that is \emph{almost} a computational basis state.

In order to mitigate this aliasing effect, we also considered so-called ``hybrid'' schedules in addition to those that are purely exponential or polynomial. A hybrid schedule interleaves a given schedule with short linear sequences in order to \emph{reduce} the likelihood that any given depth (of $\mathcal{Q}$ calls) in the schedule will be pathological as to yield a vanishingly small contribution to the Fisher information. To be more precise, given a schedule $\left\{ k_{1},k_{2},...\right\}$ that specifies sampling from $\mathcal{Q}^{k_{s}}$ applications of the Grover operator (as in Sect. \ref{MLE-QAE}), we implement instead the schedule $\left\{ k_{1},k_{1}+1,...k_{1}+j,k_{2},k_{2}+1,...k_{2}+j,...\right\} $. This ensures that even as a schedule may prescribe large increments in $k_{s}$ generally, that we nevertheless ``scan'' in the vicinity of every $k_{s}$ by small increments. In the rest of this text, unless otherwise specified, we will compute with $j=2$.

\section{Results} \label{sec:res}

In this subsection we will observe the effects of noise on QAE performance. The estimation problem we solve is as described in Sect. \ref{sec:toy}. Our primary goals are to i) Elucidate qualitatively the effects of noise, and quantitatively what performance can be expected for a given noise level and desired accuracy regime, ii) Correspondingly, show how the optimal Grover schedule depends heavily on this noise level and accuracy regime, and iii) Compare the effects of different types of noise.

What we expect to find qualitatively is that, for a given strength of noise, a Grover schedule $\{k_s\}$ that ramps up more quickly (e.g. polynomial ones like the cubic schedule) will be more rapidly degraded by noise from experiment to experiment, and the optimality of the schedule will be superseded by a slower schedule.
Nevertheless, an important outcome to note is that \emph{at a given} (relatively shallow) total Grover operators, a fast-ramping schedule like the cubic polynomial in a noisy setting can still outperform both the conservative linear and asymptotically optimal exponential schedules in the noiseless setting!

\begin{figure}[t]
	\centering
    \includegraphics[width=0.5\textwidth]{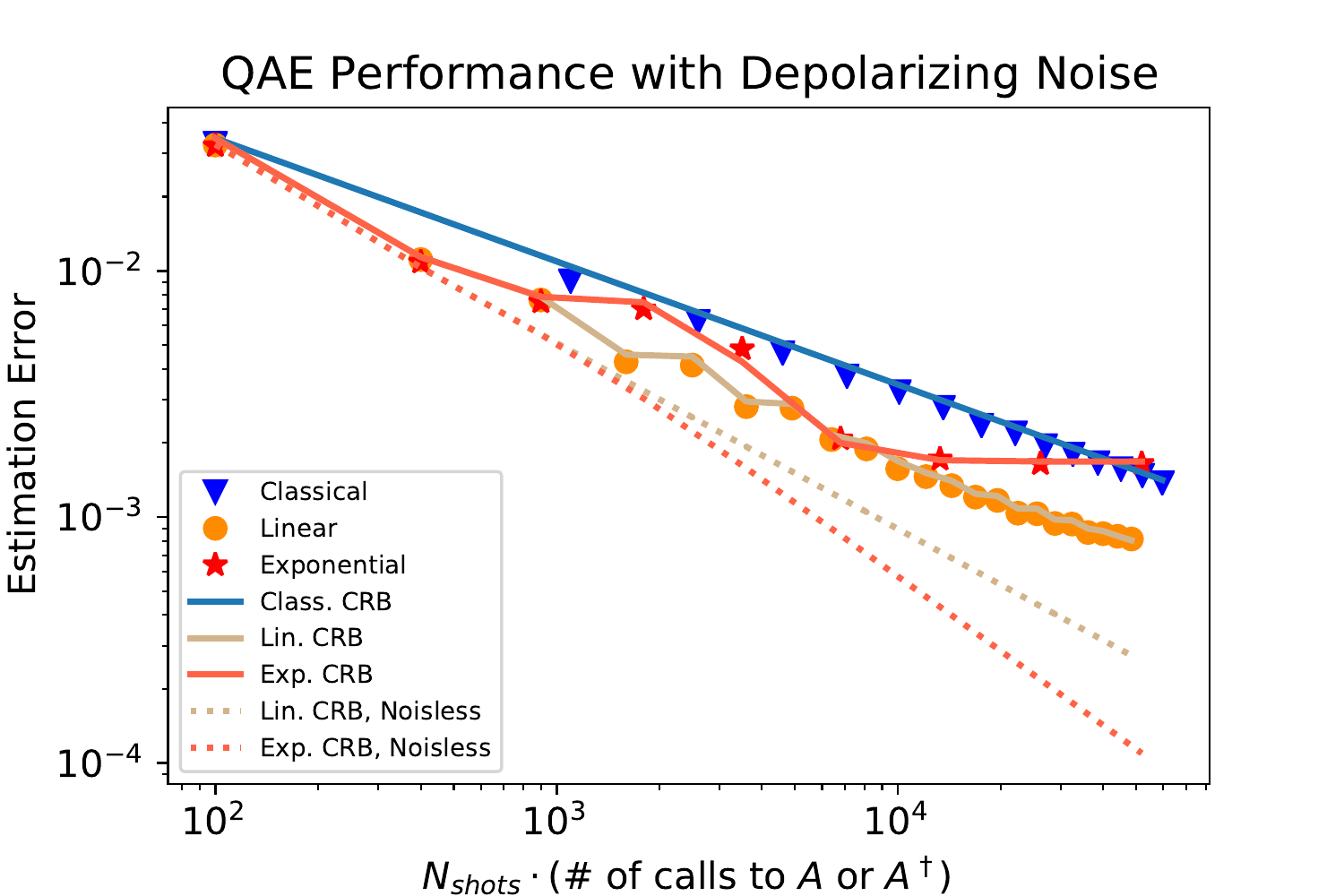}
	\caption{Effect of depolarizing noise, with parameter $\gamma = 0.05$, on linear and and exponential schedules with $N_\text{shots}=100$ shots per experiment. We plot both simulated RMS errors and the noisy CRB for each, as well as the noise-free CRBs for comparison.}
	\label{Depol}
\end{figure}

\begin{figure*} 
\begin{tabular}{ccc}
\includegraphics[width=5.6cm]{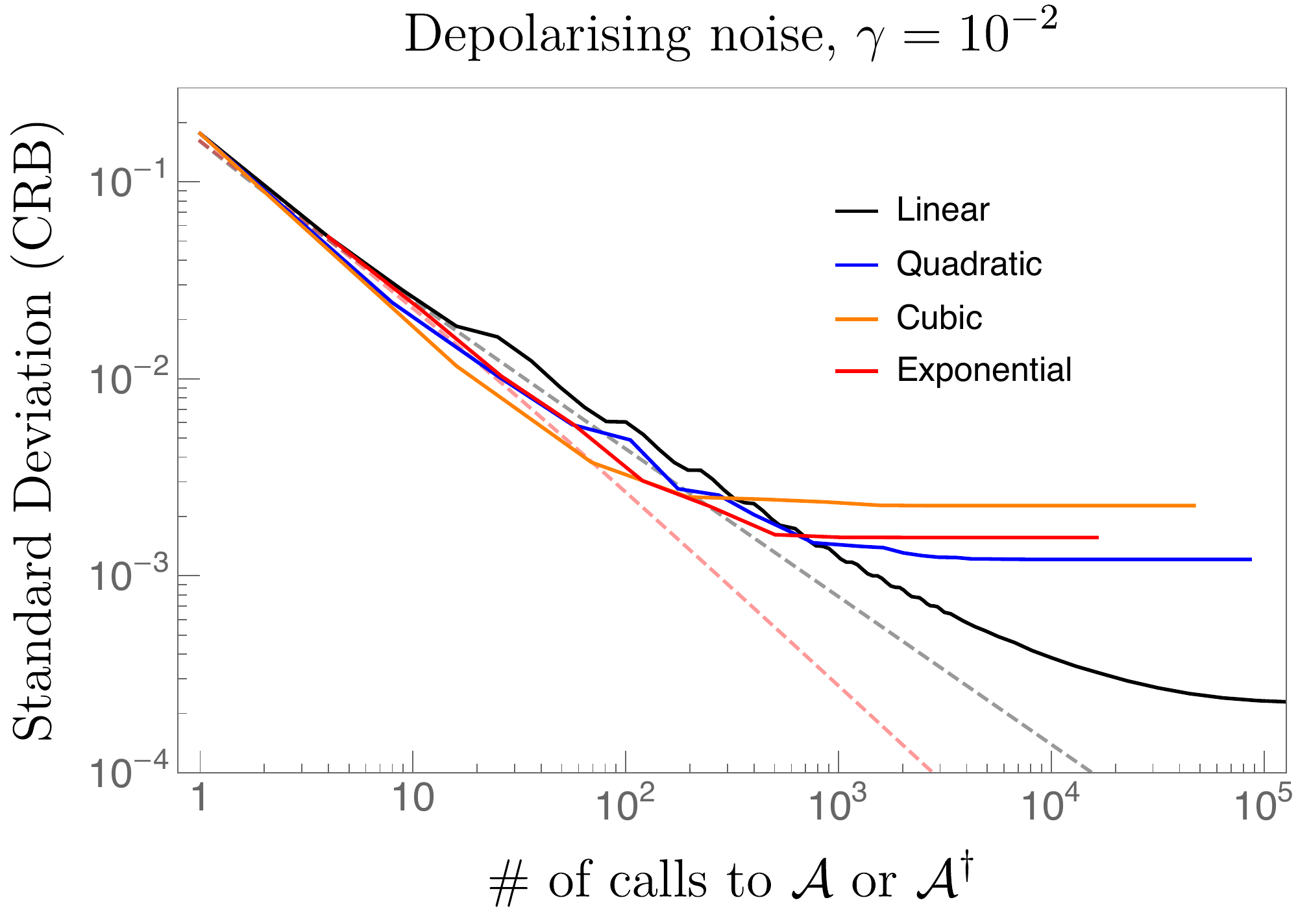} & \includegraphics[width=5.6cm]{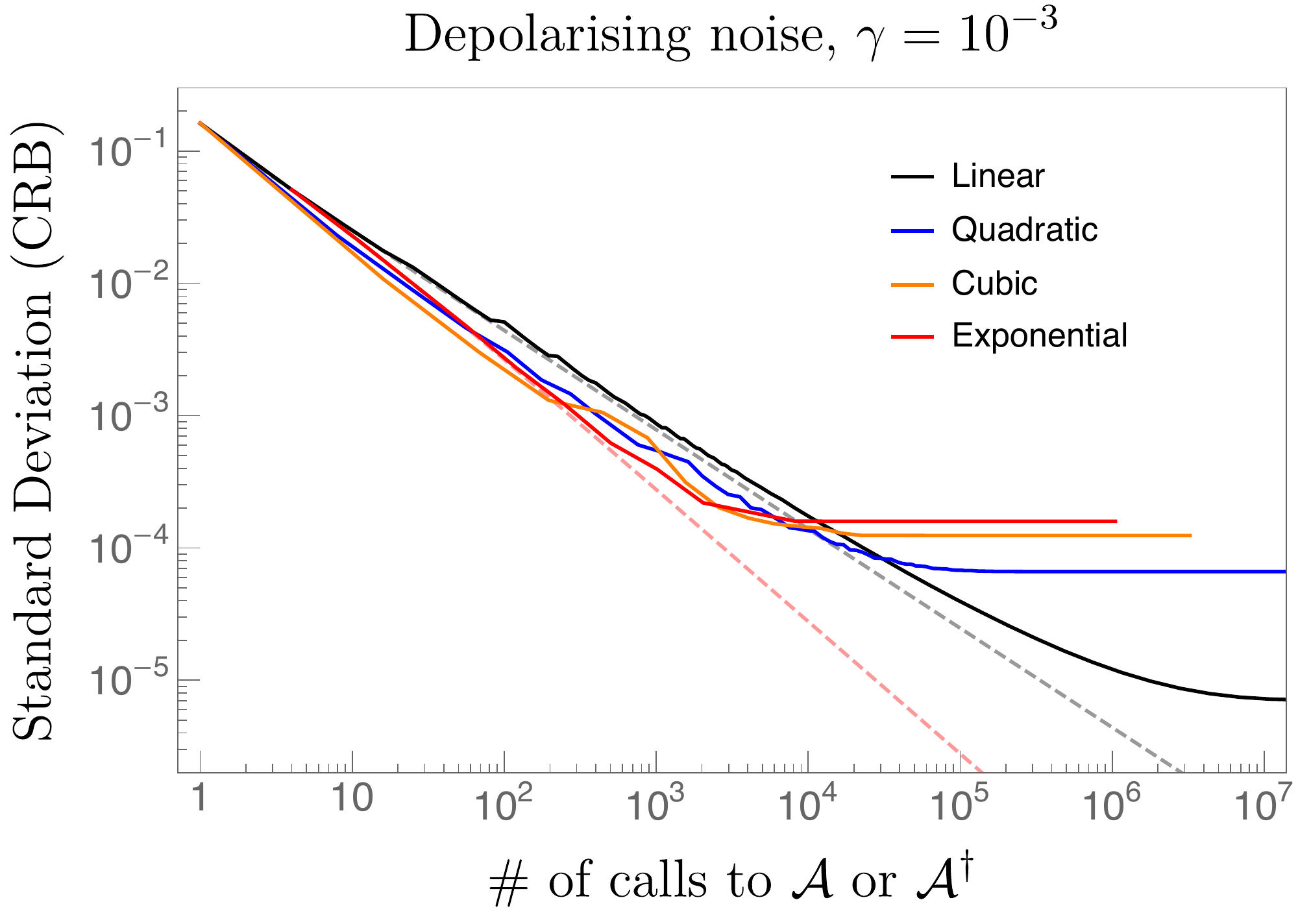} & \includegraphics[width=5.6cm]{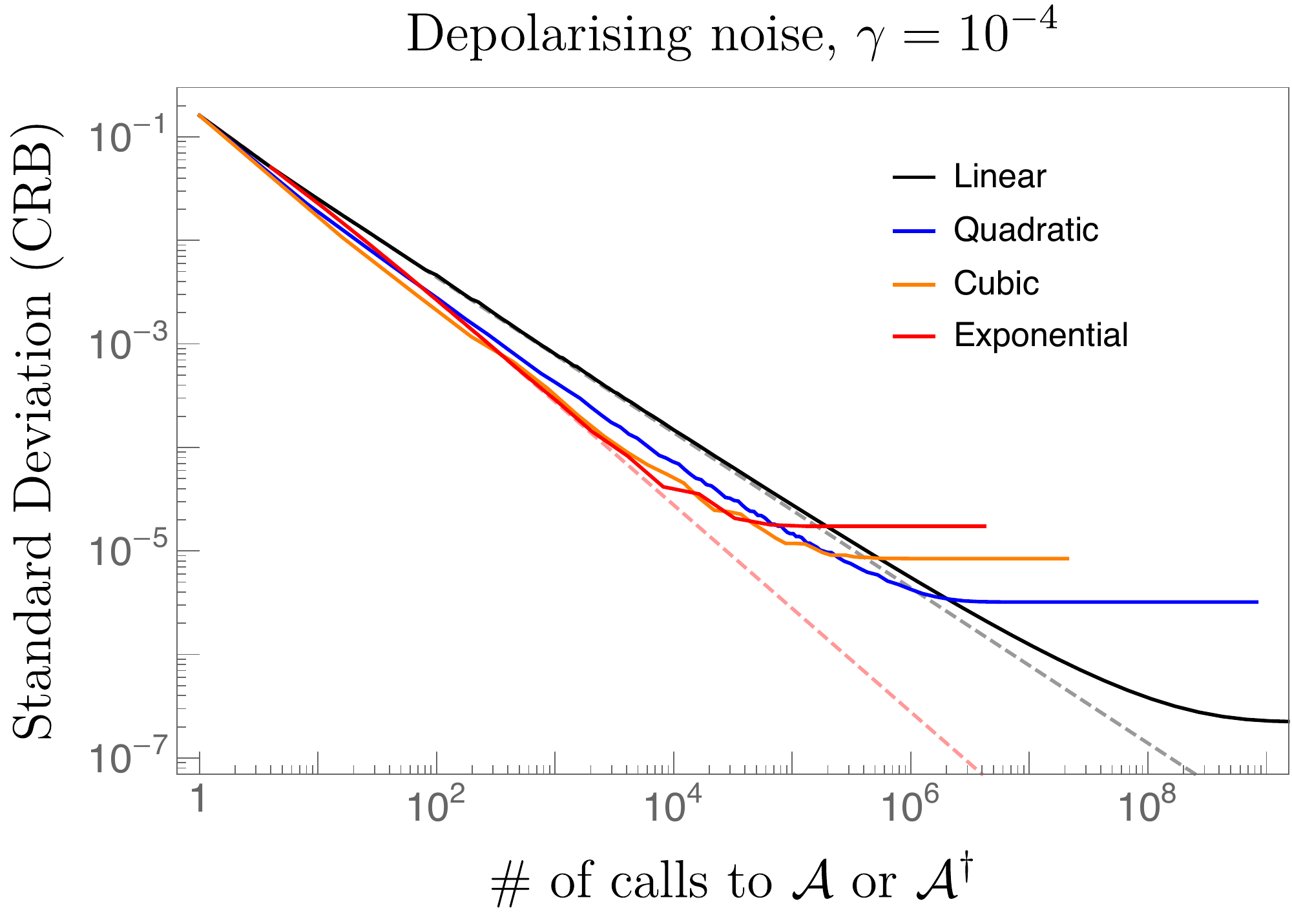}\tabularnewline
\end{tabular}

\caption{Cramer-Rao lower bounds on the Standard Deviation in estimators of amplitude $\alpha$, for various depolarising strengths and using different schedules. The dashed lines are the linear and exponential CR bound in the noise-free case, for comparison. Note that here we consider the single shot case, $N_\text{shots}=1$, and thus the x-axis must be scaled by a factor of $100$ when comparing to those of Figs. \ref{Clean}, \ref{Depol}, and \ref{optimal_schedule}.}
\label{manyDepols}
\end{figure*}

\begin{figure*} 
\begin{centering}
\begin{tabular}{ccc}
\includegraphics[width=5.5cm]{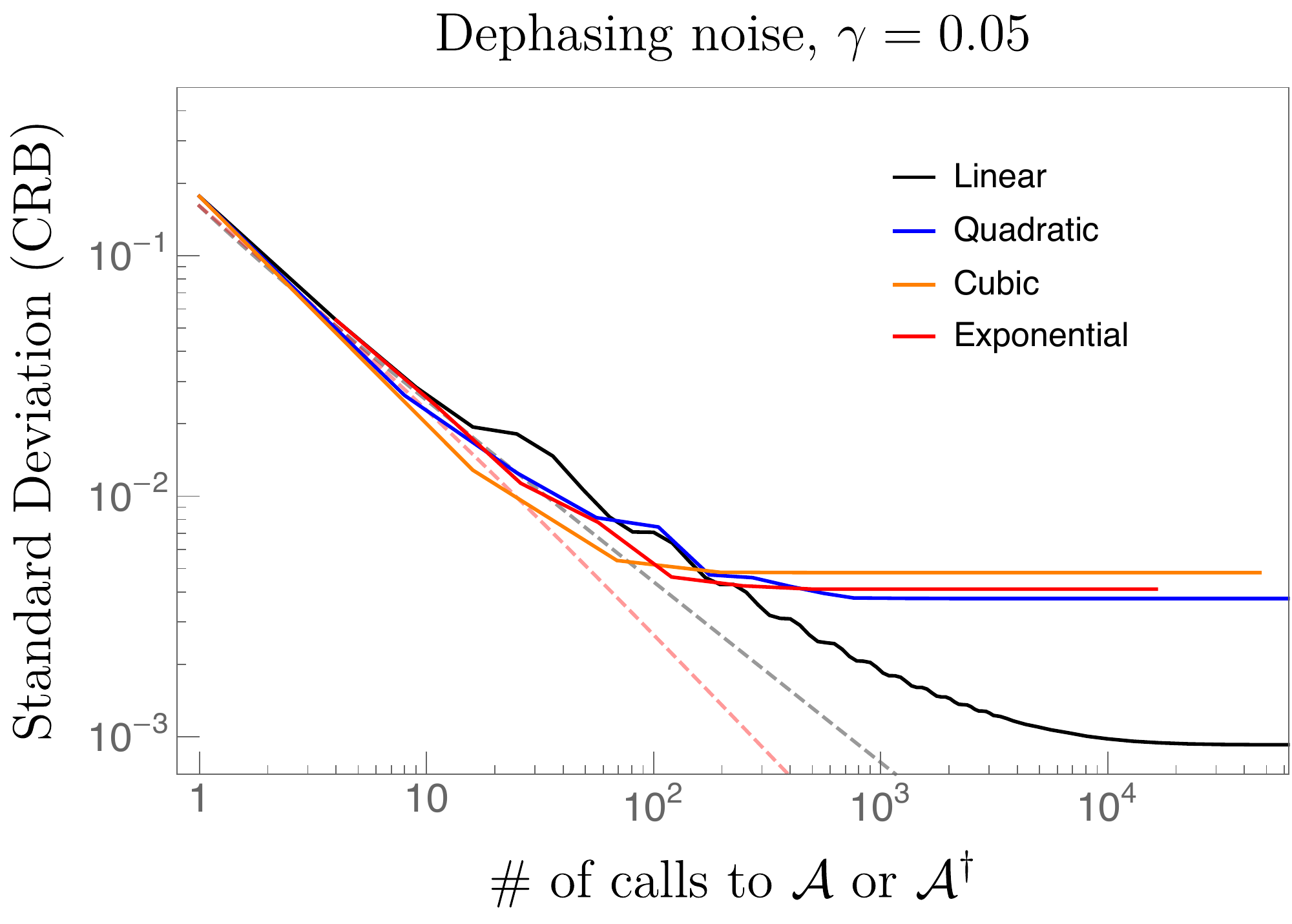} & \includegraphics[width=5.5cm]{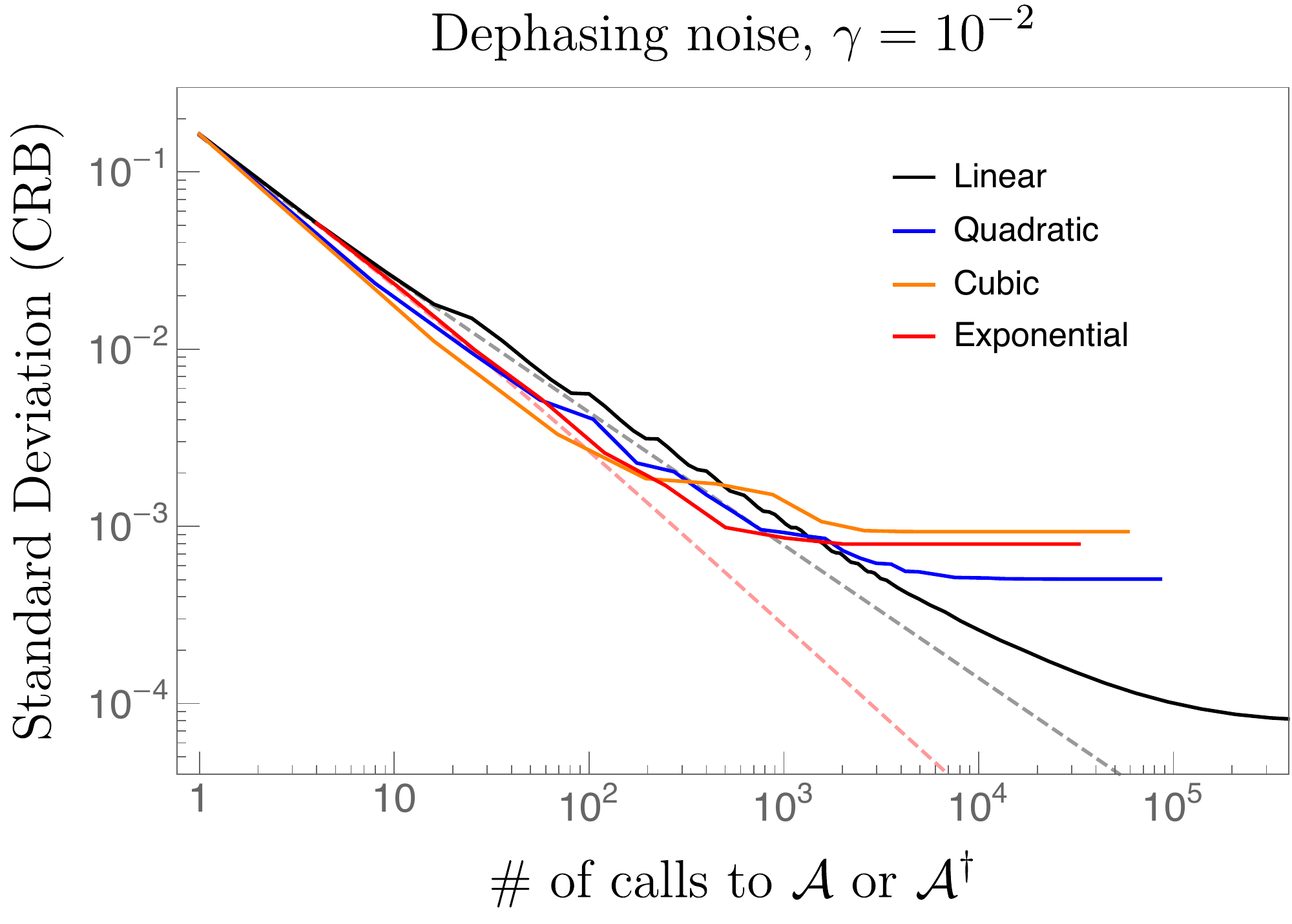} & \includegraphics[width=5.5cm]{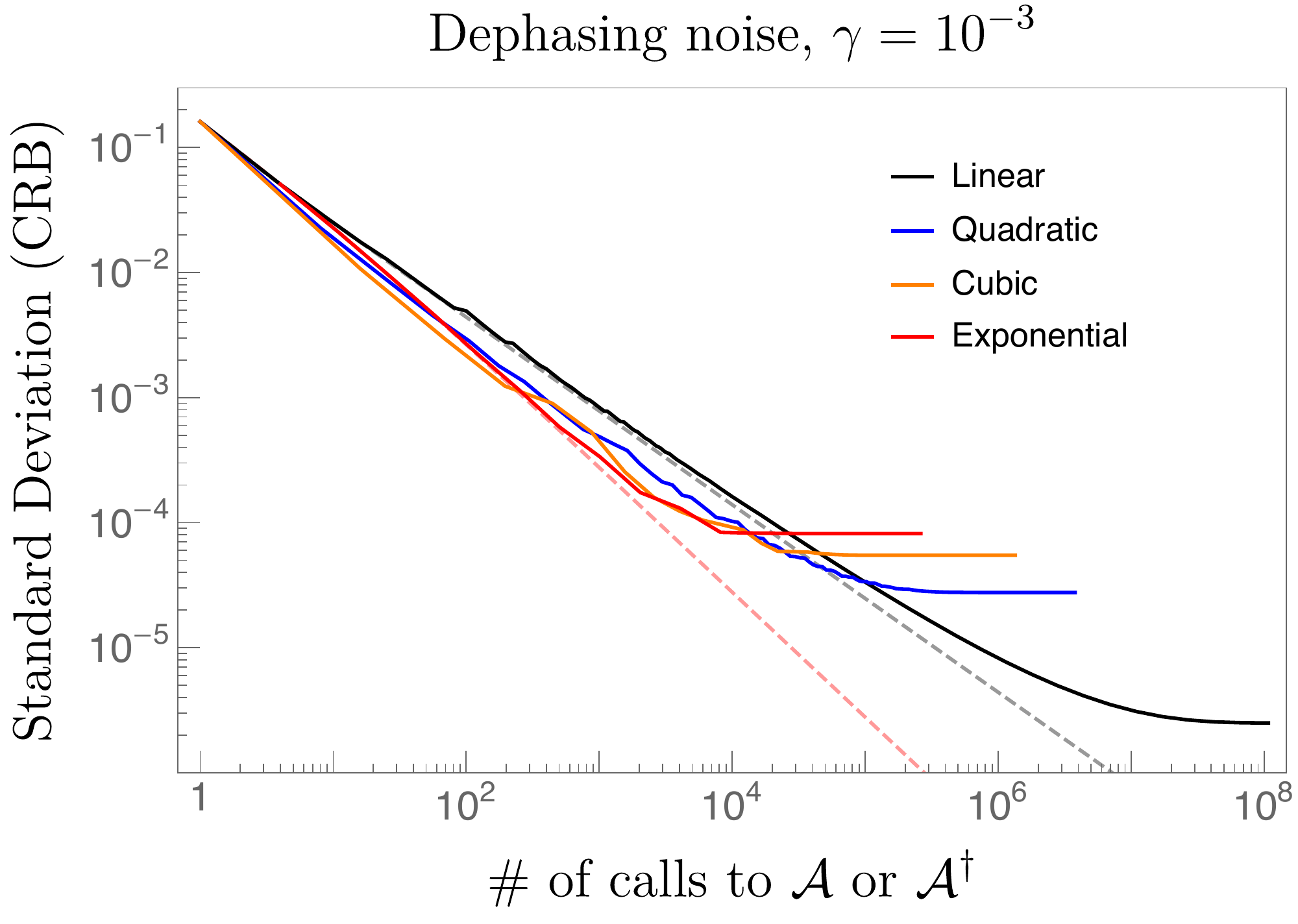}\tabularnewline
\includegraphics[width=5.5cm]{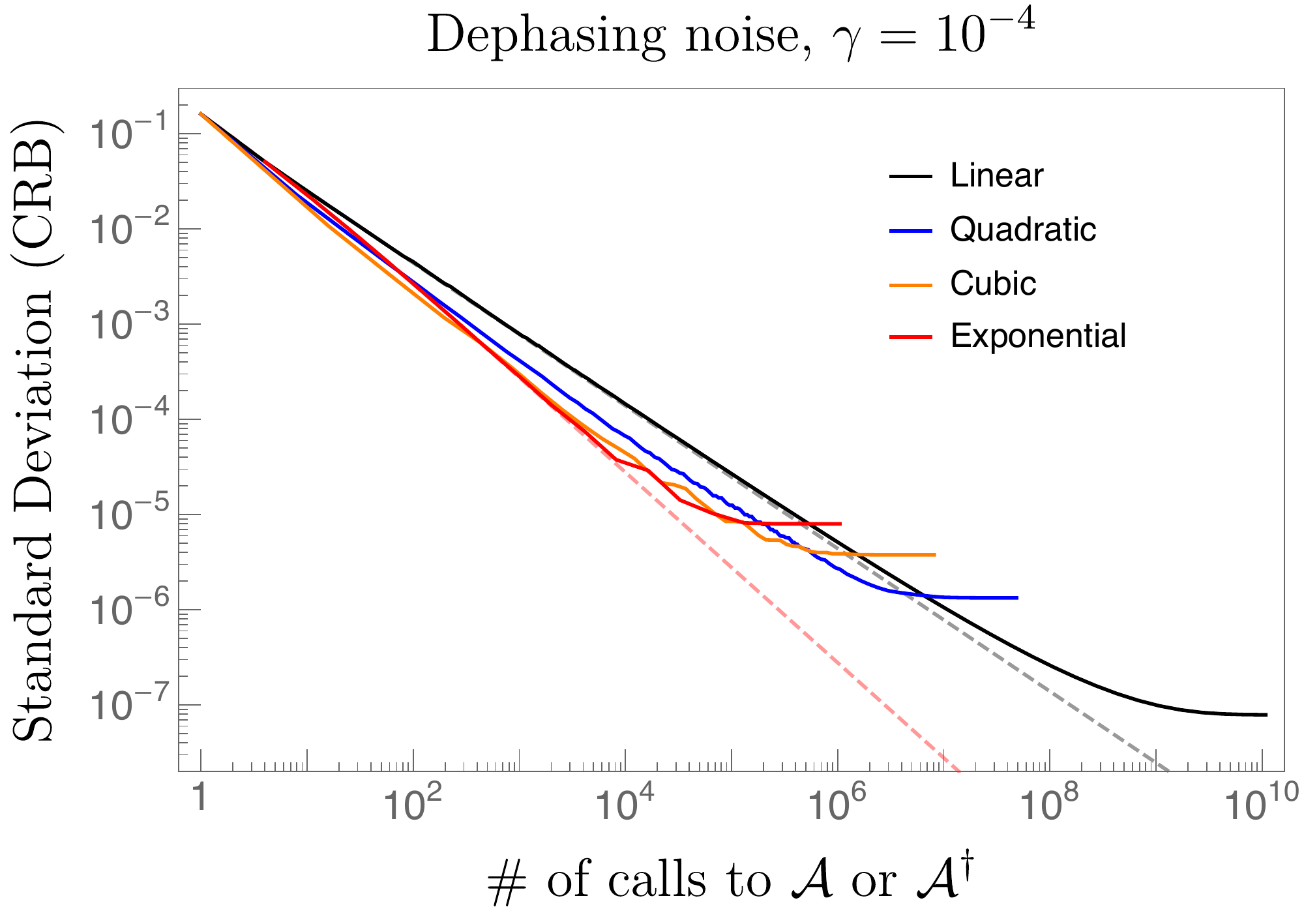} & \includegraphics[width=5.5cm]{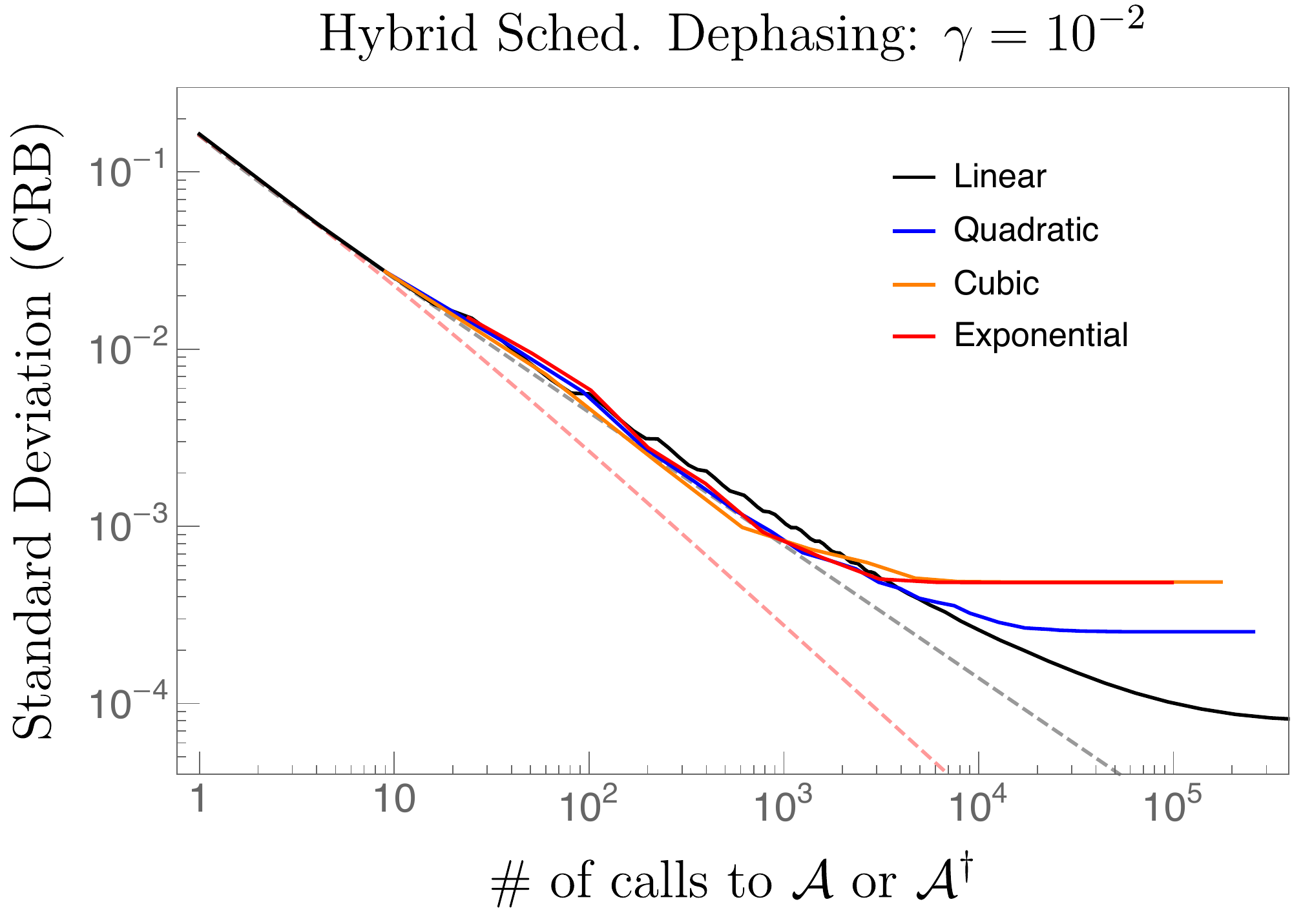} & \includegraphics[width=5.5cm]{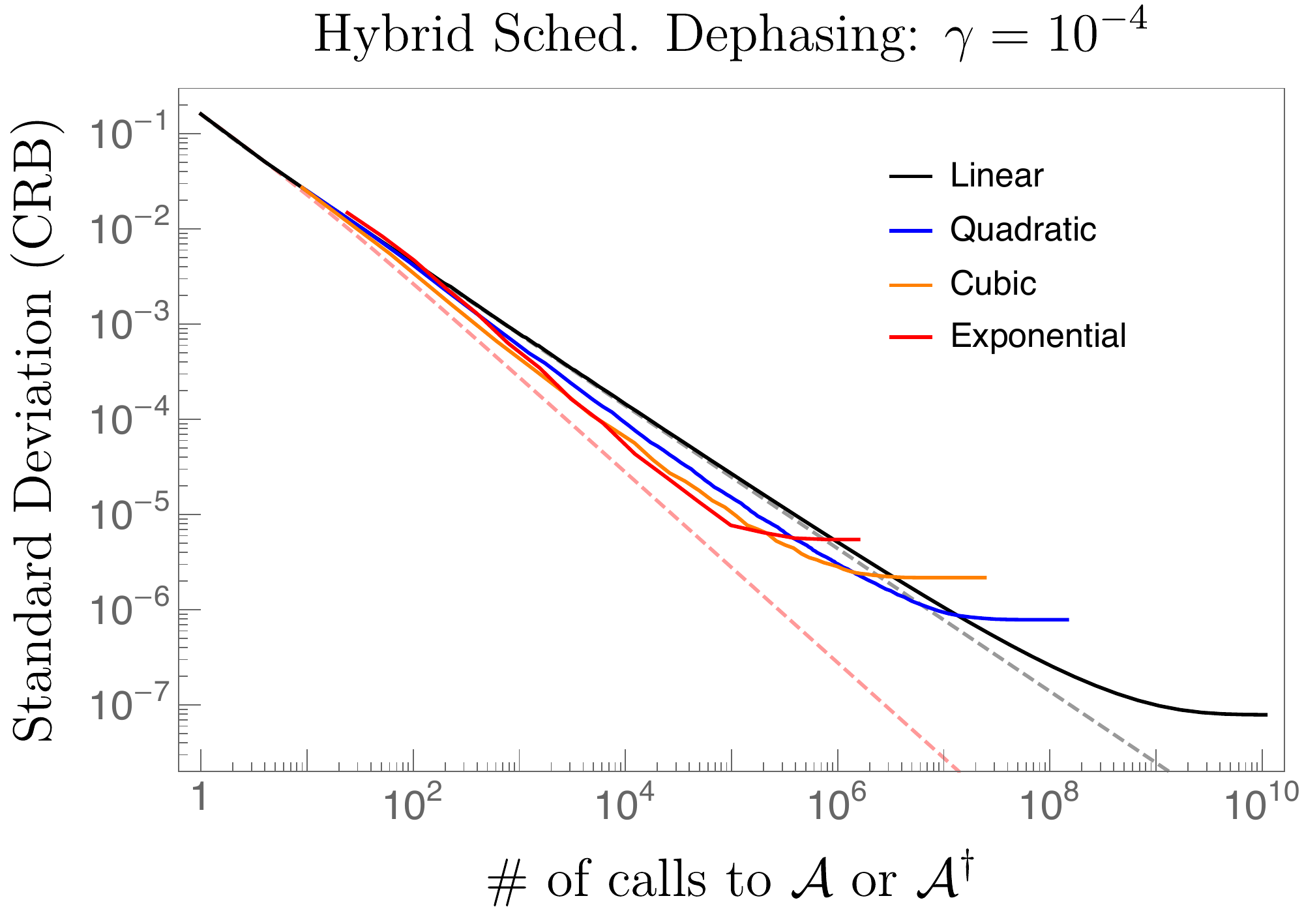}\tabularnewline
\end{tabular}
\par\end{centering}
\caption{Cramer-Rao lower bounds on the Standard Deviation in estimators of amplitude $\alpha$, for various dephasing strengths. The dashed lines are the linear and exponential CR bound in the noise-free case, for comparison. In addition to the linear, quadratic, cubic, and exponential schedules, we also show a hybrid schedule that attempts to mitigate aliasing features described in Sect \ref{aliasing}. We see that the QAE performance in the presence of dephasisng noise, for a given $\gamma$ is nearly equivalent to that in the presence of depolarizing noise. Note that here we consider the single shot case, $N_\text{shots}=1$, and thus the x-axis must be scaled by a factor of $100$ when comparing to those of Figs. \ref{Clean}, \ref{Depol}, and \ref{optimal_schedule}.}
\label{manyDephase}
\end{figure*}

\begin{figure*} 
\begin{centering}
\begin{tabular}{ccc}
\includegraphics[width=5.5cm]{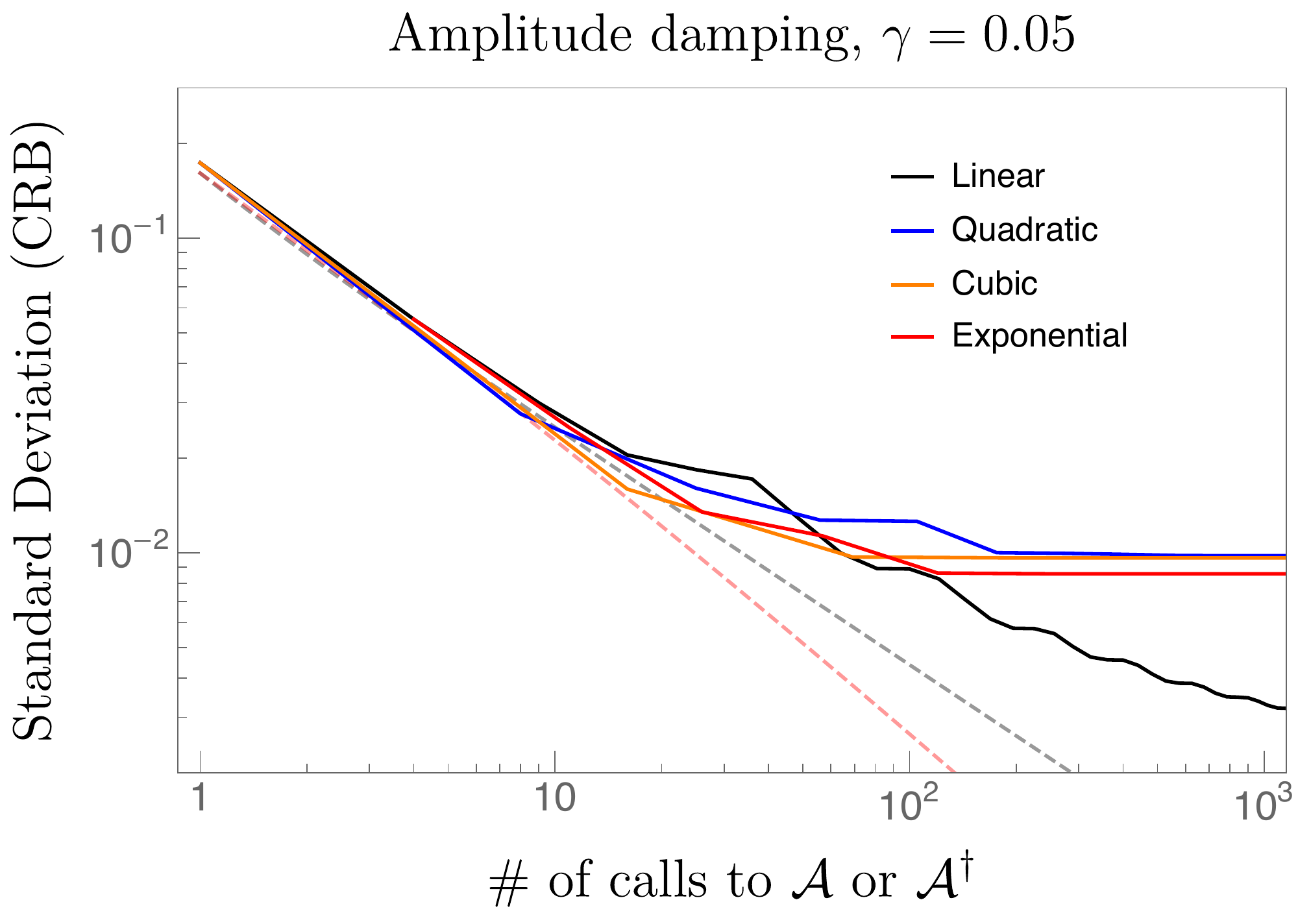} & \includegraphics[width=5.5cm]{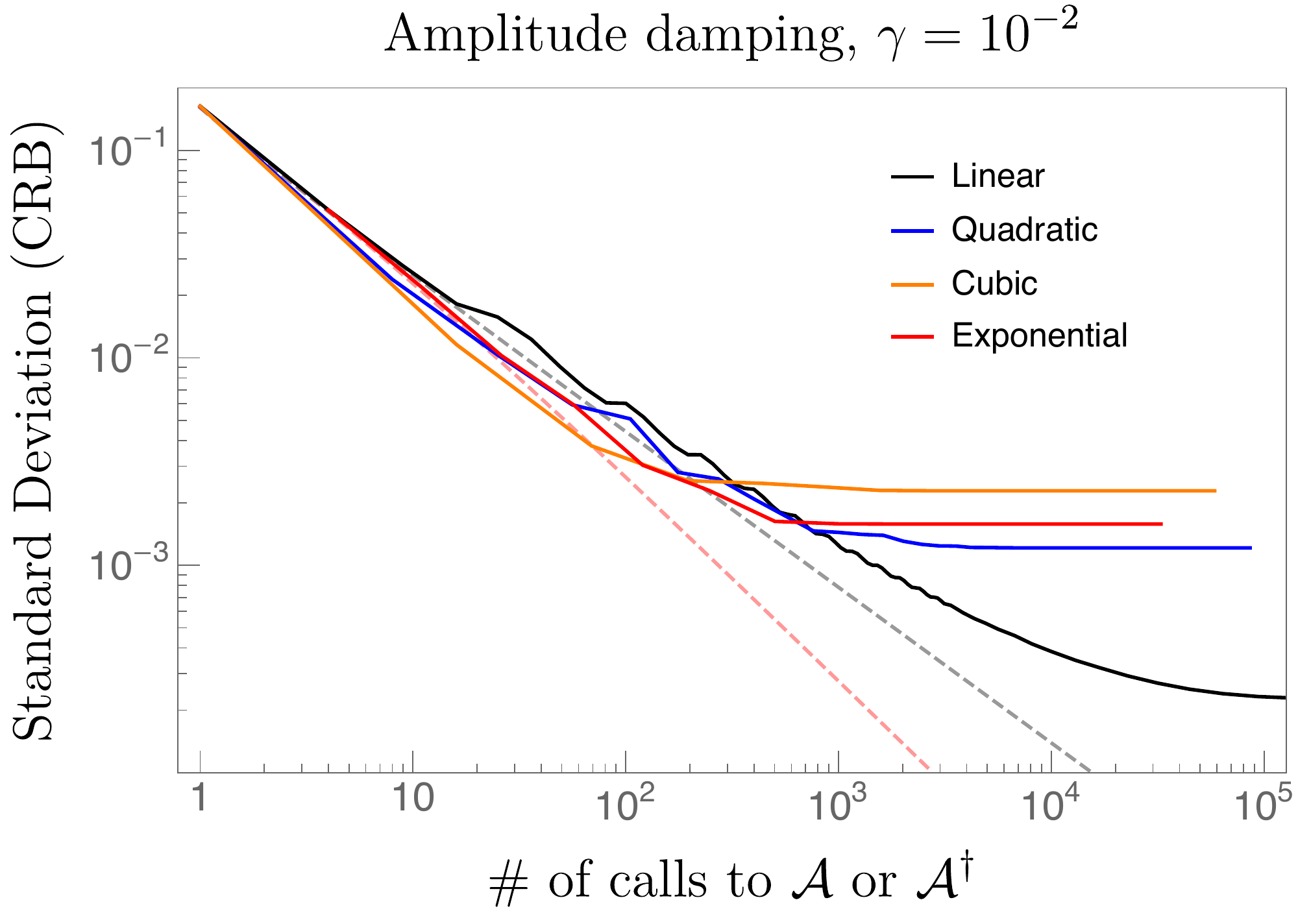} & \includegraphics[width=5.5cm]{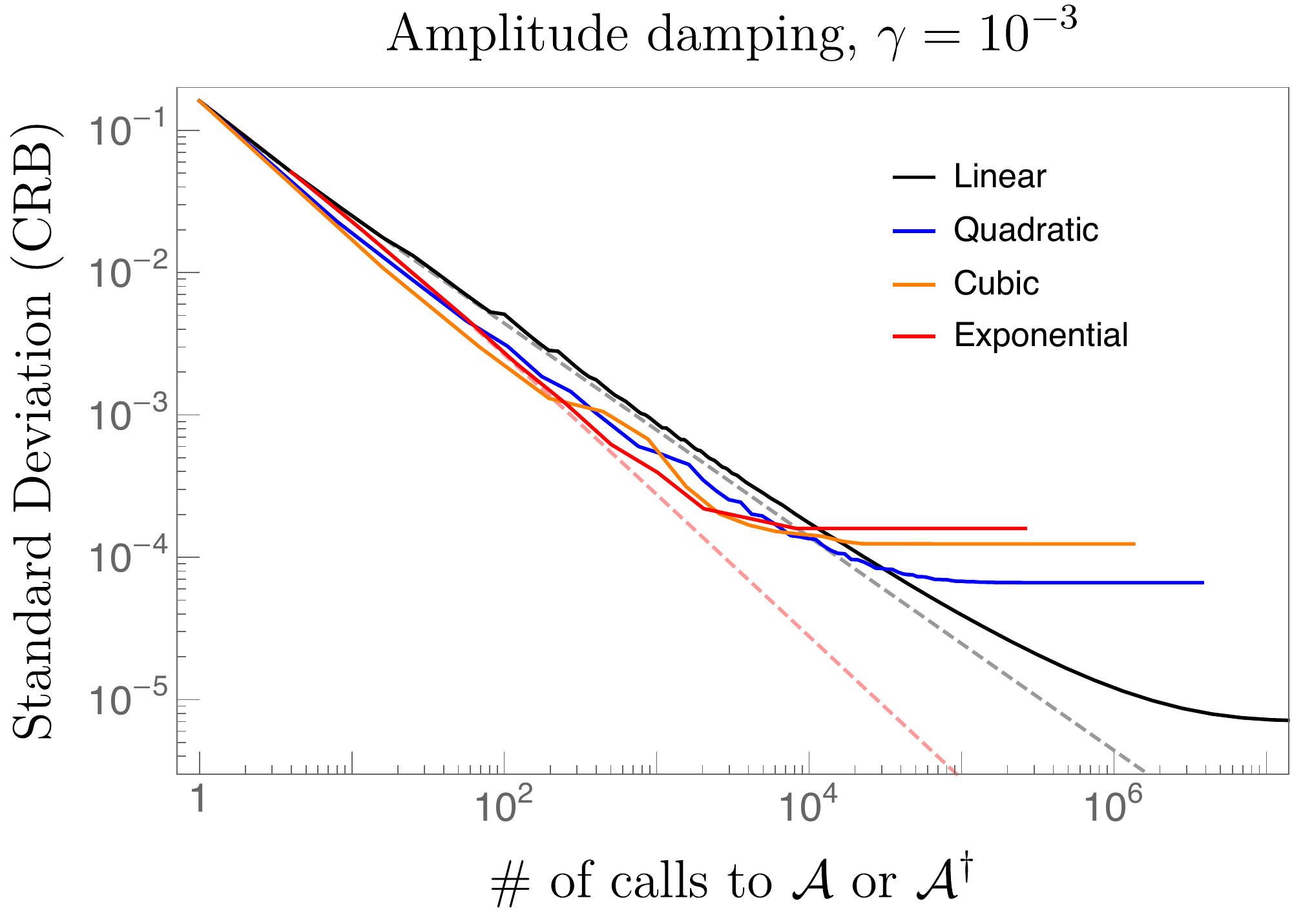}\tabularnewline
\includegraphics[width=5.5cm]{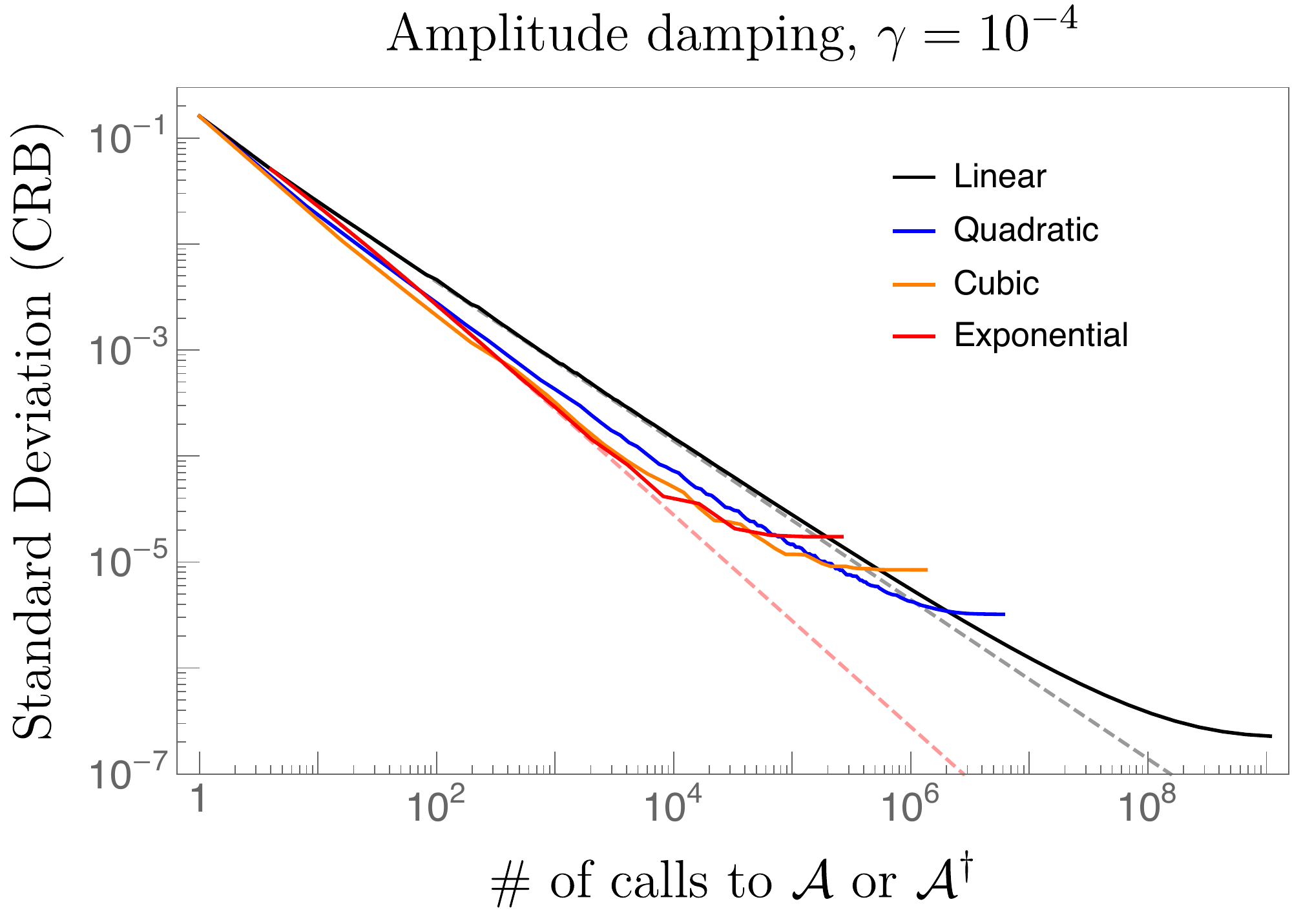} & \includegraphics[width=5.5cm]{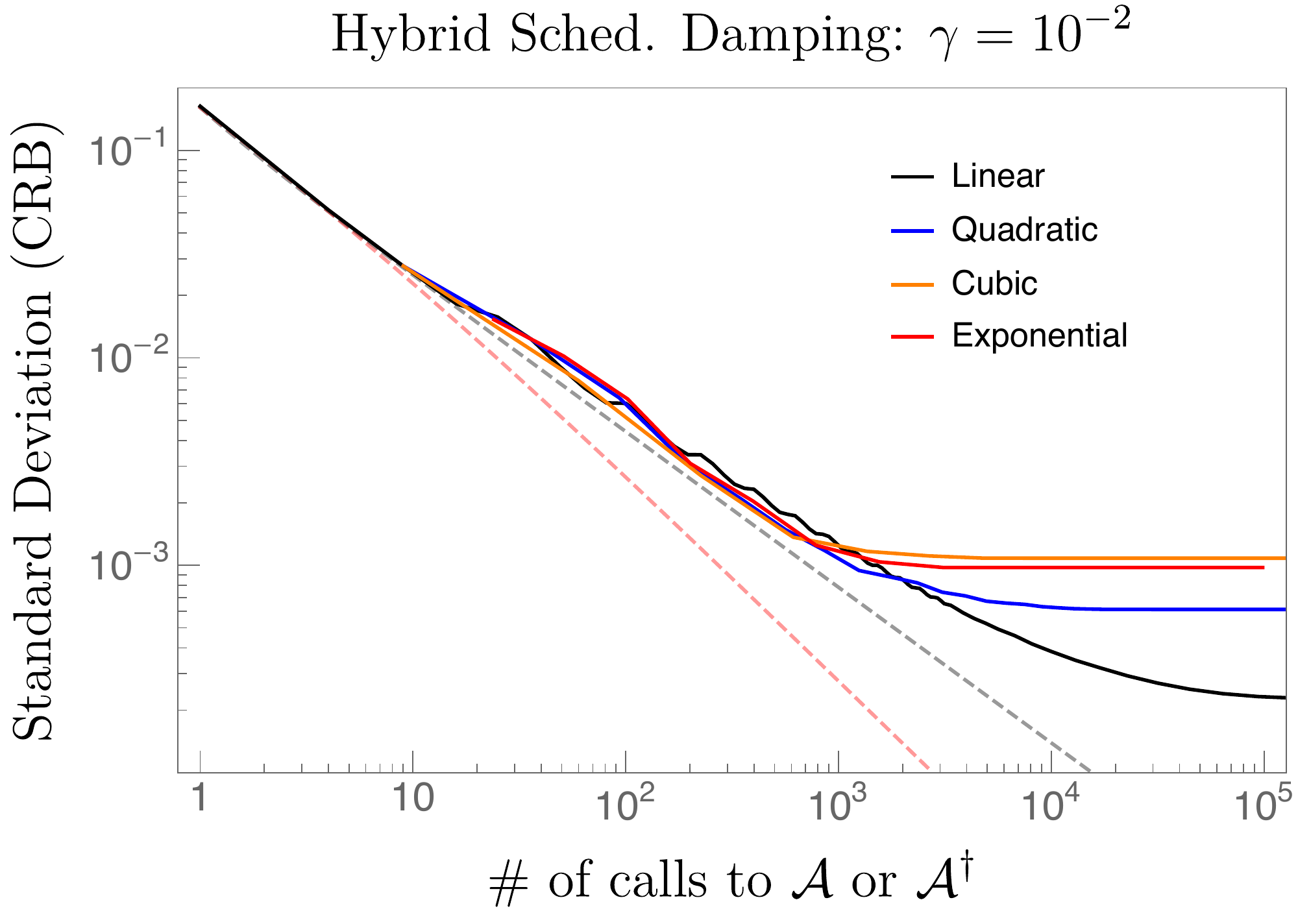} & \includegraphics[width=5.5cm]{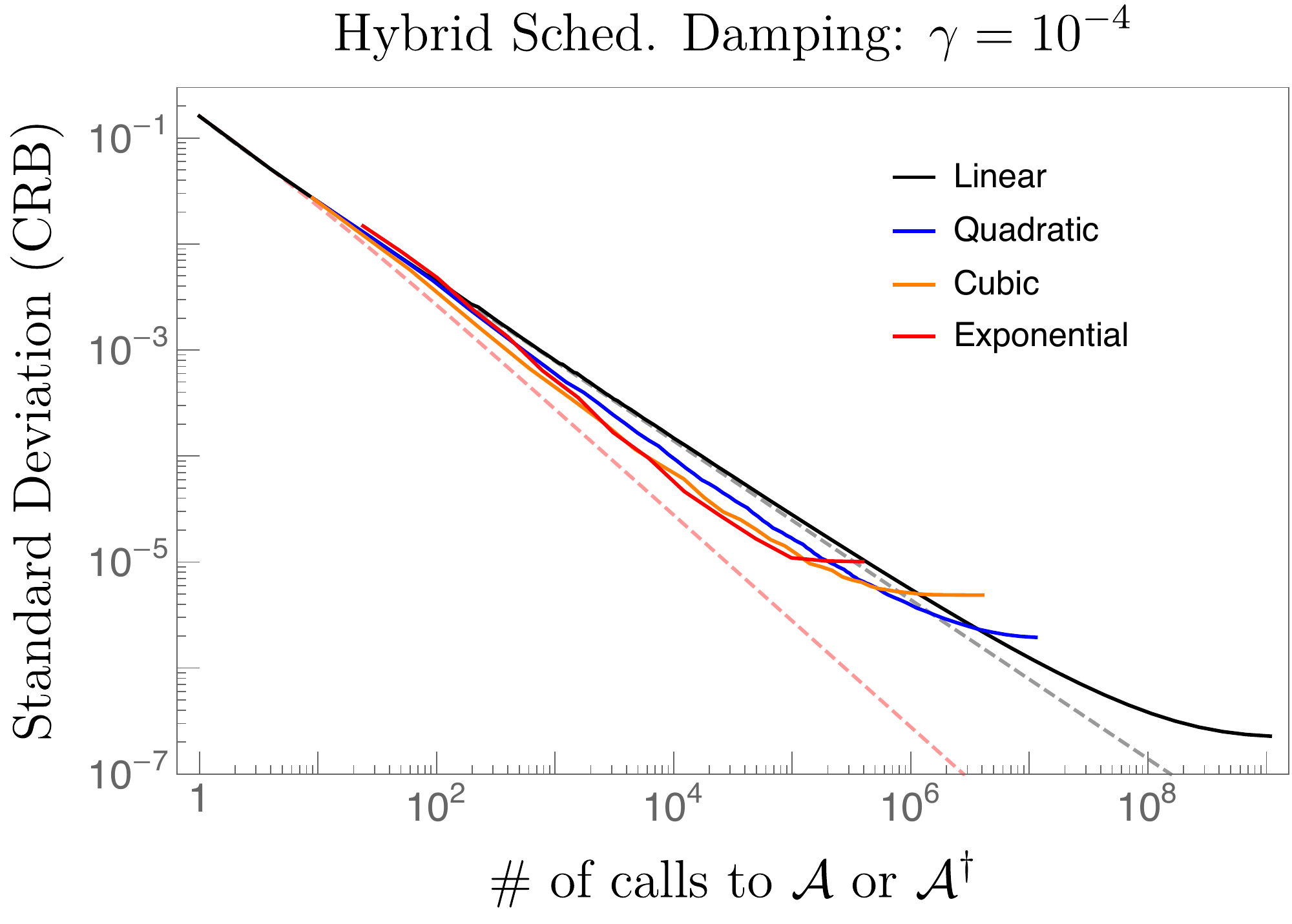}\tabularnewline
\end{tabular}
\par\end{centering}
\caption{Cramer-Rao lower bounds on the Standard Deviation in estimators of amplitude $\alpha$, for various amplitude damping strengths. The dashed lines are the linear and exponential CR bound in the noise-free case, for comparison. In addition to the linear, quadratic, cubic, and exponential schedules, we also show a hybrid schedule that attempts to mitigate aliasing features described in Sect \ref{aliasing}. In comparing with the other noise models we note that amplitude damping is generally more detrimental. Note that here we consider the single shot case, $N_\text{shots}=1$, and thus the x-axis must be scaled by a factor of $100$ when comparing to those of Figs. \ref{Clean}, \ref{Depol}, and \ref{optimal_schedule}.}
\label{manyDamping}
\end{figure*}

\begin{figure}[t]
    \centering
    \includegraphics[width=0.52\textwidth]{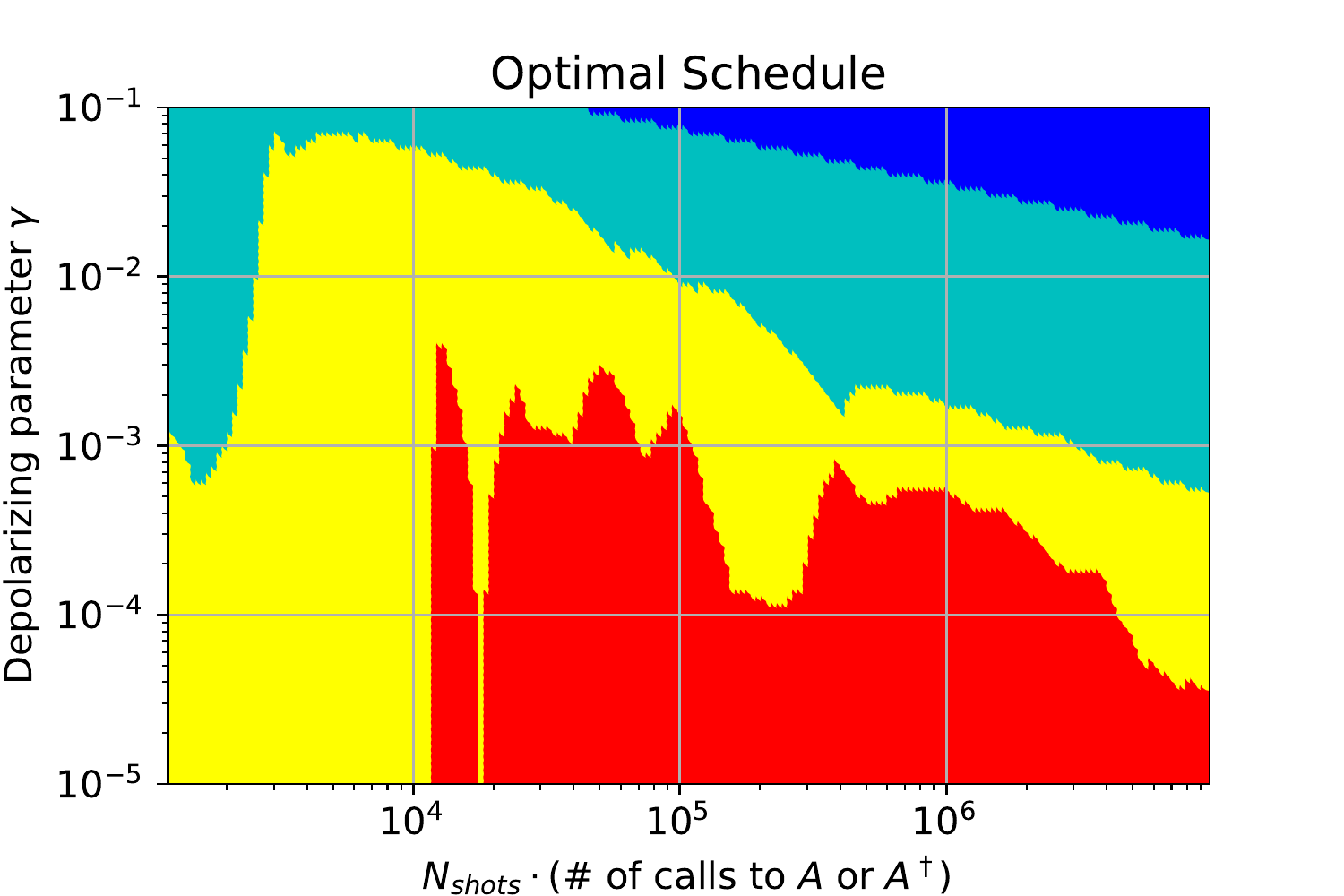}
	\caption{The optimal schedule for different total oracle calls and depolarizing noise parameters, $\gamma$. The number of shots per experiment here is $N_\text{shots}=100$. The color scheme is: \\ \emph{Dark Blue} - Classical, \\ \emph{Cyan} - Linear, \\ \emph{Yellow} - Quadratic, \\ \emph{Red} - Exponential.}
	\label{optimal_schedule}
\end{figure}

%\begin{figure}[t]
%	\centering
%    \includegraphics[width=0.50 \textwidth]{NoiseComparison.pdf}
%	\caption{Comparing the RMS error of estimation in the presence of depolarizing and dephasing noises on a linear schedule, each with parameter $\gamma = 0.1$. \eb{Will likely remove}}
%	\label{NoiseComparison}
%\end{figure}

We begin with depolarizing noise, described in Sect.~\ref{depolSect}. As an initial example, and to compare with Fig.~\ref{Clean}, let us perform QAE with linear and exponential schedules, and in the presence of depolarizing noise of strength $\gamma = 0.05$. In Fig. \ref{Depol} we plot the RMS error resulting from this and the corresponding CR bounds, computed via Eq. (\ref{noisyF}). We also plot the noise-free CR bounds for comparison. First, we note importantly that the presence of noise does not impact our being in the asymptotic regime, as the simulated MLE results show tight agreement with corresponding noisy CR bound. Second, the effect of noise is qualitatively as expected; both the linear and exponential schedules are degraded further and further by noise as each schedule proceeds, but the exponential schedule is significantly more greatly affected due to its rapidly increasing circuit depth. As such, for the same total number of calls to $\mathcal A$, the linear schedule quickly overtakes the exponential in performance.
Having satisfied ourselves that a modest number of shots per schedule entry is sufficient to saturate the CR bound in estimation performance (given an appropriately modified likelihood function), the remainder of our analyses eschew actual MLE performance, focusing instead on the CR bound.

In Figs.~\ref{manyDepols}-\ref{manyDamping} we plot the CR bound for different noise parameters, and also include quadratic and cubic schedules to compare with linear and exponential ones. We see that the quadratic schedule acts as a middle ground between the linear and exponential schedules, and that the cubic and exponential performances are often quite similar due to the similarity in their schedules. We also note that as the noise strength increases, the optimal schedule for a given number of calls to $\mathcal A$ reduces to slower schedules, with the linear schedule reliably becoming optimal as total Grover calls increases. This quality can be further seen in Fig. \ref{optimal_schedule}. In this figure we display what the optimal schedule is (achieving lowest estimation error) for a varying number of total calls and for varying depolarizing noise parameters.
% Note that for small noise and number of $\mathcal Q$ calls the quadratic schedule is more optimal than exponential; this is simply due to the fact that early in the schedules the quadratic schedule ramps up more quickly, $k_s = 0, 1, 4, 9, 16 \cdots$ than does the exponential one $k_s = 0, 1, 2, 4, 8, 16 \cdots$. The cubic schedule, similarly, ramps up even faster to begin, and a similar effect can be seen Fig. \ref{manyDepols}, where for a small number of Grover operations the \emph{noisy} cubic schedule can still slightly outperform the \emph{noise-free} exponential schedule.
Note that in all noise models, various schedules compete for optimality at different total number of calls to the Grover operator, with intermediate schedules (e.g. quadratic or cubic) becoming optimal even when compared to noise-free CR bounds for the linear and exponential schedules.
This points out the fact that when not in the (Grover) asymptotic regime the choice of optimal schedule can be far more subtle than the asymptotic analysis reveals, depending subtly on an interplay between the degree of amplification afforded by QAA and more calls to $\mathcal Q$ and system noise.

Another salient point to note here, is that unlike the depolarising and damping channels, dephasing noise of comparable magnitude affects the quality of QAE much more slowly, leaving the asymptotic error saturated at relatively smaller values.
We had briefly mentioned in Sect.~\ref{dephasingSect} that the dephasing channel stands out amongst the three noise models under consideration because it possesses a decoherence-free eigenbasis (i.e. the $\pm \hat{z}$ axis).
Therefore, as additional calls are made to Grover operator $\mathcal Q$ in the QAA procedure and the state $\ket{\psi}$ precesses, the action of dephasing noise varies between being maximally destructive (i.e. when $\ket{\psi}\approx \ket{\pm}$ to being completely trivial (i.e. when $\ket{\psi}\approx \ket{0}$ or $\ket{\psi}\approx \ket{1}$.
This points to the $T_1$ time as a more important metric when tailoring QAE for near-term quantum devices, with $T_2$ times being a ``milder'' constraint.

Finally, we note also that Figs.~\ref{manyDephase}~and~\ref{manyDamping} we explored so-called hybrid schedules described in Sect.~\ref{aliasing}.
These schedules intersperse a ``faster'' schedule (e.g. quadratic, cubic, or exponential) with linear ones.
In essence, hybrids schedules trade off absolute saturation of the CR bound in the best-case scenario for less-bad performance in the worst-case scenario.
Whereas in the worst-case scenario a faster schedule may yield a $\ket{\psi}_{\text{eff}}$ that is almost polar and therefore not yield a better estimate despite additional Grover operations, a hybrid schedule limits such scenarios to no-worse than the scaling of a comparable linear schedule.

\subsection{Discussion}

We have outlined above the results of performing MLE-QAE under different noise models and strengths. Our simulations and CR-bound calculations were performed under the circumstance that we have a complete and accurate description of the noise's effect on the final measurement amplitude, and thus we were able to noise-correct the likelihood functions used in estimation. In practice, however, such a complete description can be difficult to obtain, and any lack of accuracy in one's noise characterization will translate to a lack of accuracy in amplitude estimation. Understanding better this difficulty represents a fruitful line of future research.

Assuming perfect noise characterization, we have demonstrated that in the presence of noise one must choose their QAE schedule carefully so as to balance the trade-off between improved ideal performance and the greater accumulation of noise-induced errors. In particular our results indicate that performing an exponential schedule on a reasonably sized problem is untenable in the near future, as is achieving the Heisenberg limit. A linear schedule on the other hand, while not achieving the canonical ideal scaling, nevertheless does provide a potentially significant quantum advantage and is much more feasible to achieve on near-future hardware. In comparing the results of different noise models, we have found quantitatively similar results between depolarizing and damping noise. Dephasing noise, on the other hand, is generally less detrimental to QAE performance than comparable damping channels, pointing to the importance of ensuring that any QAE being performed be restricted to a runtime much less than the oft-quoted $T_1$ (or damping) time.

Lastly, let us take the opportunity to discuss our results in the context of realistic noise strengths in near-term devices. For simplicity we will focus on two-qubit gates, as these induce the greatest error and tend to dominate the Grover operator. In our simulated problem, Sect. \ref{sec:toy}, the number of CNOT gates on the ancilla in a single Grover operation can be shown to be $6n+4$ (three sets of controlled $R_y$ rotations between each $n$-register qubit and the ancilla, each rotation of which has two CNOTs, plus $4$ CNOTs from a Toffoli gate). Given an error rate on each CNOT, $\gamma_\text{CNOT}$, the resulting error rate per Grover operation is expected to be $\gamma \approx 1-(1-\gamma_\text{CNOT})^{6n+4}$.

In our results we have used the small number of qubits $n=3$, the resulting Grover operation of which has $22$ CNOTS acting on the ancilla qubit. Assuming a CNOT error rate of $\gamma_\text{CNOT} \sim 10^{-2}$, typical in current and near-term hardware, this will correspond to a Grover error rate of $\gamma \approx 0.2$, which from Fig. \ref{optimal_schedule} we see is beyond the point at which even a linear schedule can provide advantage over classical sampling for any reasonably significant number of calls/samples. Furthermore this will only get worse as the number of qubits $n$ increases. 

To what strength of noise must we reduce in order to have a hope of gaining quantum advantage on a reasonably sized problem? While achieving the Heisenberg bound seems unlikely without full error-correction, one may still potentially use a linear schedule to obtain a sub-Heisenberg but nevertheless significant scaling advantage over classical sampling. Let us assume that a real-world problem will need at least $n=10$ qubits ($1024$ bins to represent the sampled distribution), which in our example corresponds to $64$ CNOTs on the ancilla per Grover operation. Based on Fig. \ref{optimal_schedule}, let us conservatively estimate that $\gamma = 10^{-3}$ is the order on which a linear schedule can still provide advantage for a reasonable total number of calls and samples. The necessary noise level per CNOT is then seen to be $\gamma_\text{CNOT} \approx 10^{-5}$, a three orders of magnitude improvement over the best currently available hardware, though still potentially achievable in the NISQ era.

\section{Conclusion}   \label{sec:con}

In this work we have performed an initial examination of the effects of noise on the performance of Quantum Amplitude Estimation by simulating the algorithm in the presence of depolarizing, dephasing, and amplitude damping noise. We demonstrate that the canonical Heisenberg scaling purported by QAE is unlikely to be achievable by near-term hardware for any relevant-sized problem. However, one can improve QAE's tolerance to noise by an appropriate choice of strategy (Grover schedule), at the cost of a worse-than-ideal error scaling, such as in choosing a linear schedule over an exponential schedule. Although a linear schedule cannot achieve the ideal Heisenberg scaling, it remains a potentially significant improvement over classical sampling. We furthermore demonstrated that, at least within the simplified noise models considered, amplitude damping noise appears particularly detrimental.

A critical step in this study was the ability to exactly compute the expected error on the final measurement amplitudes (and thus the likelihood functions) resulting from our chosen noise models. Any bias that one may accidentally introduce from an incorrect error characterization becomes a bias in one's final amplitude estimation, and thus having an accurate description of the error is essential for performing QAE. For more realistic gate-noise models, however, deducing an accurate description of the error on the final amplitude itself becomes a computationally hard problem. This is on top of the problem of accurately characterizing the noise of qubits on real hardware \cite{Merkel_2013, blumekohout2013robust}.

These difficulties necessitate the further study of QAE in the presence of realistic noise models and of the optimal algorithmic strategies for mitigating the effects of noise on estimation accuracy. For example a further work could perform a full study on noisy QAE when one isn't able to exactly characterize the noise model, and examine how best to mitigate the resulting bias in one's estimate.

\section*{Acknowledgements and Contributions}

Simulation results in this study were performed using IBM Quantum's Qiskit SDK.
The authors would like to extend thanks to Tomoki Tanaka, Shumpei Uno, Yohichi Suzuki, and Rudy Raymond, who wrote the Qiskit community tutorial~\cite{githubrepo} from which the code used in this study was derived.
The authors further thank J. Kanem for useful discussions, and E. MacGowan for useful feedback.
In this work, E.B. and W.K.T. performed research and numerical calculations while O.G. assisted in the writing of this manuscript.

\appendix

\section{The Likelihood and Fisher Information} \label{A1}

Consider a probabilistic model $p_{\theta}(h)$ over a random variable $H$. The model is parameterized by $\theta$, such that $p_\theta(h)$ is the probability of the model sampling $H=h$ given parameter $\theta$. Then the corresponding Likelihood function given outcome $h$ is a function of $\theta$ and is defined as exactly this model probability:
\begin{align}
    \mathcal{L}(\theta | h) \equiv p_\theta(H = h).
\end{align}
The method of Maximum-Likelihood Estimation (from which many machine learning algorithms can be derived) is based on maximizing (usually numerically) the Likelihood-function, namely finding the parameter $\tilde \theta$ such that
\begin{align}
\tilde \theta = \text{argmax}_\theta \, \mathcal{L}(\theta | h),
\end{align}
for a given sample (or set of samples) $h$. This answers the question of: given an observed data $h$, what parameter of my model $\theta$ should I choose such that my model best fits the data? In practice one will often instead use the \emph{log-Likelihood} $\log \mathcal{L}(\theta |h)$ as it is easier to optimize.

Now, to discuss the Fisher Information, and to conform with the main text, let us instead consider the likelihood as a function of the amplitude $\alpha = \cos^2 \theta$ instead, $\mathcal{L}(\alpha |h)$. The Fisher Information is then defined as
\begin{align} \label{F1-2}
    \mathcal{I}_f(\alpha) = \mathbb{E} \left[\left(\frac{\partial}{\partial \alpha} \ln \mathcal{L}(\alpha | h)\right)^2 \right],
\end{align}
where the expectation is taken over $p_\theta(h)$. This quantity can be understood as a way of measuring how much information a random variable $h$ has about an underlying model parameter $\alpha$. I.e. it is a measure of how impactful $\alpha$ (and thus $\theta$) is on the statistics of $h$. 

The Fisher Information is often used in conjunction with the Cramer-Rao inequality, which states that when estimating via samples from $p_a(h)$, one's estimate $\alpha'$ has a squared error that is lower bounded as 
\begin{align}
    \mathbb{E}[(\alpha' - \alpha)^2] \geq \frac{(1+ b'(\alpha))^2}{\mathcal{F}(\alpha)} + b(\alpha)^2,
\end{align}
where $b(\alpha) = \mathbb{E}(\alpha' - \alpha)$ represents any bias present in the estimate, and $b'(\alpha)$ is the bias' derivative with respect to $\alpha$. In the main text we achieve a near-zero bias by simply taking sufficiently many shots per experiment, and in this case the inequality can be show to saturate in the asypmtotic limit,
\begin{align}
    \mathbb{E}[(\alpha'-\alpha)^2] \rightarrow \frac{1}{\mathcal{F}(\alpha)}.
\end{align}
We utilize this in the main text to provide easy, analytical estimates of the performance of different Grover schedules in QAE with and without the presence noise.

\section{Fisher Information with Noise}    \label{B1}

Here we give an overview of the calculation leading to Eq. (\ref{noisyF}), the Fisher Information accounting for \emph{depolarizing} noise. In the main text we consider the result over a schedule $\{k_s\}$ of different number of Grover operators. Here we will simply compute the single-experiment, single-shot Fisher Information, as from this we can trivially arrive at the scheduled version Eq. (\ref{noisyF}) by using additivity of the Fisher information for multiple independent measurement outcomes and experiments.

Given a number of Grover operators $k$, the noisy amplitude as measured on the ancilla qubit is given by
\begin{align} 
    \widetilde{\alpha}_k = \gamma_k \alpha_k + \frac{1}{2}(1-\gamma_k),
\end{align}
where $1-\gamma_k$ is the probability of error and $\alpha_k = \cos^2 ((2k+1)\theta)$ is the corresponding noiseless amplitude. Given a fraction $p_k$ of ``good" measurements, the noisy likelihood function per shot in this experiment is thus
\begin{align}
    \widetilde{\mathcal{L}}_k = \tilde{\alpha}_k^{p_k}(1-\tilde{\alpha}_k)^{1-p_k}.
\end{align}
The noisy Fisher Information is then
\begin{align} \label{F2}
    \mathcal{\tilde I}_f^{(k)}(\alpha) &= \mathbb{E} \left[\left(\frac{\partial}{\partial \alpha} \ln \tilde{\mathcal{L}}_k\right)^2 \right],
\end{align}
where the derivative is taken with respect to $\alpha = \cos^2 \theta$. To compute the expectation, recall that $p_k$ follows a Binomial distribution, the statistics of which (in the presence of noise) are
\begin{align} \label{noisyBin}
    \mathbb{E}(p_k) &= \tilde{\alpha}_k, \nonumber \\
    \mathbb{E}(p_k^2) &= \text{var}(p_k) + \mathbb{E}(p_k)^2 = \tilde{\alpha}_k.
\end{align}

Our task is first to compute the following:
\begin{align}
    \frac{\partial}{\partial \alpha} \ln \tilde{\mathcal{L}}_k &= \left(\frac{p_k}{\tilde{\alpha}_k} - \frac{1-p_k}{1-\tilde{\alpha}_k}\right)\frac{\partial \tilde{\alpha}_k}{\partial \alpha} \\
    &= \sqrt{\frac{\alpha_k(1-\alpha_k)}{\alpha(1-\alpha)}} \frac{\gamma_k(2k+1)}{\tilde{\alpha}_k(1-\tilde{\alpha}_k)}(p_k-\tilde{\alpha_k})
\end{align}
where in the second line we have used the simple-to-confirm equality
\begin{align}
    \frac{\partial \tilde \alpha_k}{\partial \alpha} = \gamma_k \frac{\partial \alpha_k}{\partial \alpha}
    = p_k \sqrt{\frac{\alpha_k (1-\alpha_k)}{\alpha(1-\alpha)}} (2k+1).
\end{align}

From here, the Fisher Information is given by
\begin{align}
    \mathcal{\tilde I}_f^{(k)}(\alpha) &= \frac{\alpha_k(1-\alpha_k)}{\alpha(1-\alpha)}\frac{\gamma_k^2 (2k+1)^2}{\tilde{\alpha}_k^2(1-\tilde{\alpha_k})^2}\mathbb{E}((p_k-\tilde{\alpha_k})^2) \nonumber \\
    &= \frac{1}{\alpha(1-\alpha)}\gamma_k^2 \frac{\alpha_k(1-\alpha_k)}{\tilde{\alpha}_k(1-\tilde{\alpha_k})}(2k+1)^2,
\end{align}
where in the second line we used the Binomial statistics as in Eq. (\ref{noisyBin}).

In the case of performing a schedule $\{k_s\}$ of experiments, each of which is performed for $N_s$ shots, then by additivity of the Fisher Information it is straightforward to confirm that the Fisher Information for the schedule as a whole is given by
\begin{align}
    \mathcal{\tilde I}_f(\alpha) = \frac{1}{\alpha(1-\alpha)}\sum_s \gamma_{k_s}^2 \, \frac{\alpha_{k_s}(1-\alpha_{k_s})}{\tilde{\alpha}_{k_s} (1- \tilde{\alpha}_{k_s})} \cdot N_s (2k_s+1)^2.
\end{align}

\section{Computing trajectory of input state under repeated Grover operations and noisy channels.}\label{A3}
\begin{figure}
\begin{centering}
\includegraphics[width=6cm]{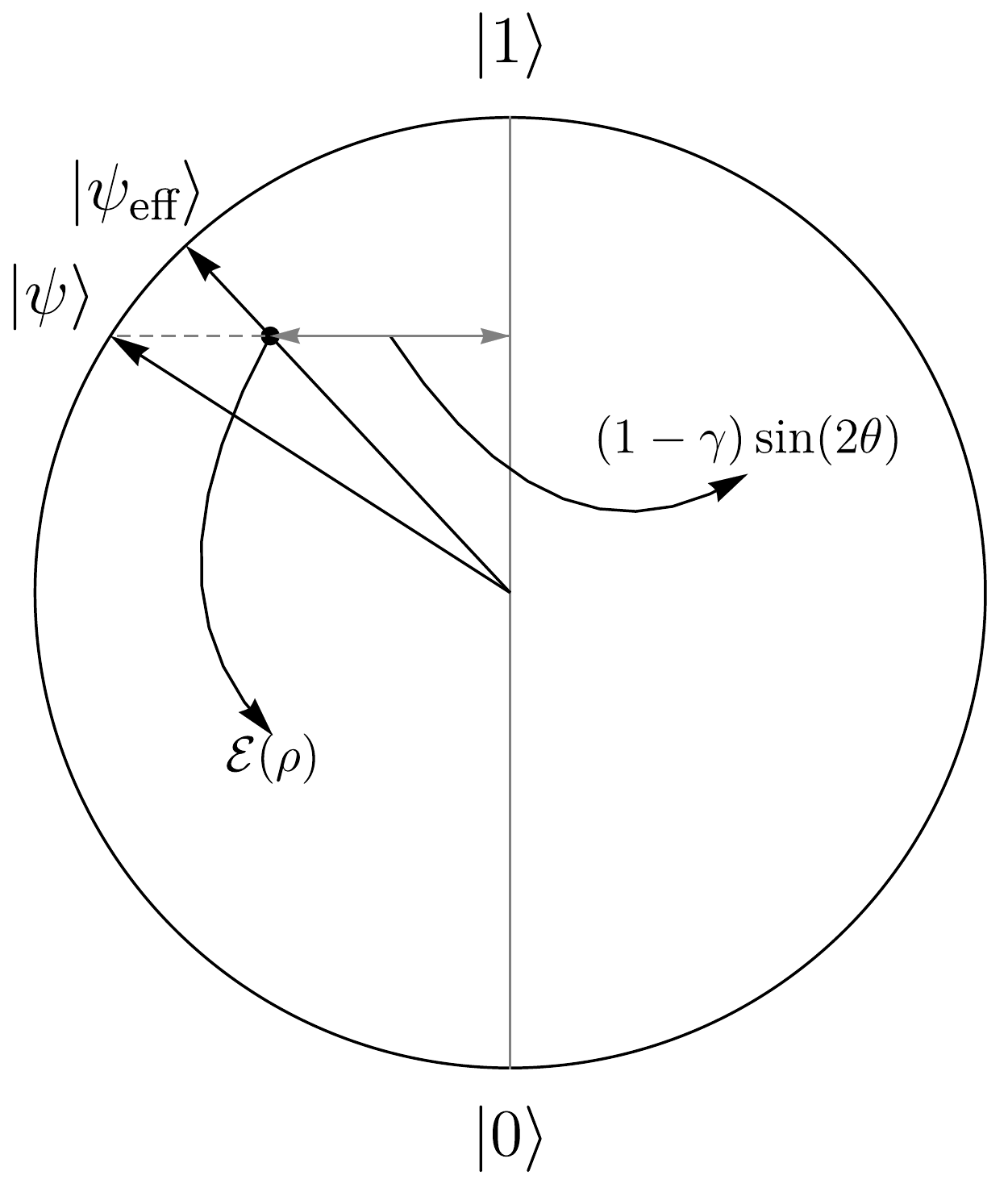}
\par\end{centering}
\caption{Illustration of the effect of dephasing noise on $\left|\psi\right>=\cos\theta\left|0\right>+\sin\theta\left|1\right>$.\label{fig:BlochBlob}}
\end{figure}
Here we briefly describe the process by which, given an initial $\theta$ (or equivalently, an amplitude $\alpha=\cos^2\theta$), one can efficiently numerically compute the trajectory of $\varphi$ (or equivalently $\left|\psi_{\text{eff}}\right>=U^{(\mathcal{E},m)}\left|\psi\right>=\cos\varphi\left|0\right>+\sin\varphi\left|1\right>$). In other words, we wish to compute the action of some \emph{effective} unitary transformation (as defined in the decomposition described in Section~\ref{sec:CRB}) on the initial state $\left|\psi\right>=\cos\theta\left|0\right>+\sin\theta\left|1\right>$, under a sequence of Grover operator ($\mathcal{Q}$) and noise channels ($\mathcal{E}$). While the method described here does not provide a closed form description of the effects of $\mathcal{E}$ on the QAE procedure it nevertheless allows one to compute, numerically, quantities of interest like the Fisher information or the likelihood function when performing MLE.

For pedagogical reasons, let us suppose we are analysing the dephasing channel as described in Eq.~\ref{eq:defDephasing}. Because we are considering the dephasing channel acting only on the readout qubit (a single-qubit register; see Fig.~\ref{circuit}), we can geometrically picture this as a contraction along every direction of the Bloch sphere except for the $\hat{z}$ direction (i.e. the computational basis). We illustrate this in Fig.~\ref{fig:BlochBlob}. Expressing the output state $\rho_{\text{eff}}=\mathcal{E}(\left|\psi\right>\left<\psi\right|)$ as a mixture of a pure state $\left|\psi_{\text{eff}}\right>=\cos\varphi\left|0\right>+\sin\varphi\left|1\right>$ and the maximally mixed state (see Section~\ref{sec:CRB}), simple trigonometry as can be seen in Fig.~\ref{fig:BlochBlob} allows us to quickly deduce $\left|\psi_{\text{eff}}\right>$ as follows:
\begin{align}
\text{Tr}\rho_{\text{eff}}\sigma_{x} & =p_{\text{eff}}\sin2\varphi=(1-\gamma)\sin(2\theta) \nonumber \\
\text{Tr}\rho_{\text{eff}}\sigma_{z} & =p_{\text{eff}}\cos2\varphi=\cos(2\theta)
\end{align}
\[
\implies p_{\text{eff}}=\sqrt{(1-\gamma)^{2}\sin^{2}2\theta+\cos^{2}2\theta}
\]
\begin{align*}
\implies\text{Tr}\left|\psi_{\text{eff}}\right>\left<\psi_{\text{eff}}\right|\sigma_{x} & =\cos2\varphi=\frac{\cos2\theta}{p_{\text{eff}}}\\
\text{Tr}\left|\psi_{\text{eff}}\right>\left<\psi_{\text{eff}}\right|\sigma_{z} & =(1-\gamma)\sin2\varphi=\frac{\sin2\theta}{p_{\text{eff}}}
\end{align*}
Ergo, for a schedule that prescribes $m$ applications of the Grover operator ($\mathcal{Q}$) interspersed with the dephasing channel, we easily compute $\left|\psi_{\text{eff}}\right>$ and $p_{\text{eff}}$ with the following simple procedure.
\begin{algorithm}
    \begin{itemize}
    \item We are given $\theta$, $m$, and $\gamma$.
    \item Assign $\varphi_{1}=\theta$ and $p_{\text{eff}}^{(1)}=1$.
    \item \textbf{For $2\leq k\leq m$:}
    
    \uline{Dephasing channel; compute:}
    
    $p\leftarrow\sqrt{\cos^{2}2\varphi_{k-1}+(1-\gamma)^{2}\sin^{2}2\varphi_{k-1}}$
    
    $\sin2\varphi_{k}\leftarrow(1-\gamma)\sin2\varphi_{k-1}/p$
    
    $\cos2\varphi_{k}\leftarrow\cos2\varphi_{k-1}/p$
    
    $p_{\text{eff}}^{(k)}\leftarrow p_{\text{eff}}^{(k-1)}\times p$
    
    \uline{then for Grover operation; assign:}
    
    $\sin2\varphi_{k}\leftarrow\sin(2\varphi_{k}+4\theta)=\sin2\varphi_{k}\cos4\theta+\cos2\varphi_{k}\sin4\theta$
    
    $\cos2\varphi_{k}\leftarrow\cos(2\varphi_{k}+4\theta)=\cos2\varphi_{k}\cos4\theta-\sin2\varphi_{k}\sin4\theta$
    \end{itemize}
    \caption{Compute $\left|\psi_{\text{eff}}\right>$ and $p_{\text{eff}}$ under $m$ Grover operators and dephasing channels.\label{alg:propagateDephasing}}
\end{algorithm}

Now while Algorithm~\ref{alg:propagateDephasing} was constructed for a dephasing channel, it can easily be adapted for any other noisy channel $\mathcal E$ with the same geometrical arguments.

% \newline
\bibliography{main.bib}

\begin{thebibliography}{29}%
\makeatletter
\providecommand \@ifxundefined [1]{%
 \@ifx{#1\undefined}
}%
\providecommand \@ifnum [1]{%
 \ifnum #1\expandafter \@firstoftwo
 \else \expandafter \@secondoftwo
 \fi
}%
\providecommand \@ifx [1]{%
 \ifx #1\expandafter \@firstoftwo
 \else \expandafter \@secondoftwo
 \fi
}%
\providecommand \natexlab [1]{#1}%
\providecommand \enquote  [1]{``#1''}%
\providecommand \bibnamefont  [1]{#1}%
\providecommand \bibfnamefont [1]{#1}%
\providecommand \citenamefont [1]{#1}%
\providecommand \href@noop [0]{\@secondoftwo}%
\providecommand \href [0]{\begingroup \@sanitize@url \@href}%
\providecommand \@href[1]{\@@startlink{#1}\@@href}%
\providecommand \@@href[1]{\endgroup#1\@@endlink}%
\providecommand \@sanitize@url [0]{\catcode `\\12\catcode `\$12\catcode
  `\&12\catcode `\#12\catcode `\^12\catcode `\_12\catcode `\%12\relax}%
\providecommand \@@startlink[1]{}%
\providecommand \@@endlink[0]{}%
\providecommand \url  [0]{\begingroup\@sanitize@url \@url }%
\providecommand \@url [1]{\endgroup\@href {#1}{\urlprefix }}%
\providecommand \urlprefix  [0]{URL }%
\providecommand \Eprint [0]{\href }%
\providecommand \doibase [0]{http://dx.doi.org/}%
\providecommand \selectlanguage [0]{\@gobble}%
\providecommand \bibinfo  [0]{\@secondoftwo}%
\providecommand \bibfield  [0]{\@secondoftwo}%
\providecommand \translation [1]{[#1]}%
\providecommand \BibitemOpen [0]{}%
\providecommand \bibitemStop [0]{}%
\providecommand \bibitemNoStop [0]{.\EOS\space}%
\providecommand \EOS [0]{\spacefactor3000\relax}%
\providecommand \BibitemShut  [1]{\csname bibitem#1\endcsname}%
\let\auto@bib@innerbib\@empty
%</preamble>
\bibitem [{\citenamefont {Shor}(1999)}]{Shor_1999}%
  \BibitemOpen
  \bibfield  {author} {\bibinfo {author} {\bibfnamefont {Peter~W.}\
  \bibnamefont {Shor}},\ }\bibfield  {title} {\enquote {\bibinfo {title}
  {Polynomial-time algorithms for prime factorization and discrete logarithms
  on a quantum computer},}\ }\href {\doibase 10.1137/S0036144598347011}
  {\bibfield  {journal} {\bibinfo  {journal} {SIAM Review}\ }\textbf {\bibinfo
  {volume} {41}},\ \bibinfo {pages} {303--332} (\bibinfo {year} {1999})},\
  \Eprint {http://arxiv.org/abs/https://doi.org/10.1137/S0036144598347011}
  {https://doi.org/10.1137/S0036144598347011} \BibitemShut {NoStop}%
\bibitem [{\citenamefont {Grover}(1997)}]{Grover1997}%
  \BibitemOpen
  \bibfield  {author} {\bibinfo {author} {\bibfnamefont {Lov~K.}\ \bibnamefont
  {Grover}},\ }\bibfield  {title} {\enquote {\bibinfo {title} {Quantum
  mechanics helps in searching for a needle in a haystack},}\ }\href {\doibase
  10.1103/PhysRevLett.79.325} {\bibfield  {journal} {\bibinfo  {journal} {Phys.
  Rev. Lett.}\ }\textbf {\bibinfo {volume} {79}},\ \bibinfo {pages} {325--328}
  (\bibinfo {year} {1997})}\BibitemShut {NoStop}%
\bibitem [{\citenamefont {McArdle}\ \emph {et~al.}(2020)\citenamefont
  {McArdle}, \citenamefont {Endo}, \citenamefont {Aspuru-Guzik}, \citenamefont
  {Benjamin},\ and\ \citenamefont {Yuan}}]{McArdle_2020}%
  \BibitemOpen
  \bibfield  {author} {\bibinfo {author} {\bibfnamefont {Sam}\ \bibnamefont
  {McArdle}}, \bibinfo {author} {\bibfnamefont {Suguru}\ \bibnamefont {Endo}},
  \bibinfo {author} {\bibfnamefont {Al\'an}\ \bibnamefont {Aspuru-Guzik}},
  \bibinfo {author} {\bibfnamefont {Simon~C.}\ \bibnamefont {Benjamin}}, \ and\
  \bibinfo {author} {\bibfnamefont {Xiao}\ \bibnamefont {Yuan}},\ }\bibfield
  {title} {\enquote {\bibinfo {title} {Quantum computational chemistry},}\
  }\href {\doibase 10.1103/RevModPhys.92.015003} {\bibfield  {journal}
  {\bibinfo  {journal} {Rev. Mod. Phys.}\ }\textbf {\bibinfo {volume} {92}},\
  \bibinfo {pages} {015003} (\bibinfo {year} {2020})}\BibitemShut {NoStop}%
\bibitem [{\citenamefont {Ryabinkin}\ \emph {et~al.}(2018)\citenamefont
  {Ryabinkin}, \citenamefont {Yen}, \citenamefont {Genin},\ and\ \citenamefont
  {Izmaylov}}]{ryabinkin_2018}%
  \BibitemOpen
  \bibfield  {author} {\bibinfo {author} {\bibfnamefont {Ilya~G.}\ \bibnamefont
  {Ryabinkin}}, \bibinfo {author} {\bibfnamefont {Tzu-Ching}\ \bibnamefont
  {Yen}}, \bibinfo {author} {\bibfnamefont {Scott~N.}\ \bibnamefont {Genin}}, \
  and\ \bibinfo {author} {\bibfnamefont {Artur~F.}\ \bibnamefont {Izmaylov}},\
  }\href@noop {} {\enquote {\bibinfo {title} {Qubit coupled-cluster method: A
  systematic approach to quantum chemistry on a quantum computer},}\ }
  (\bibinfo {year} {2018}),\ \Eprint {http://arxiv.org/abs/1809.03827}
  {arXiv:1809.03827 [quant-ph]} \BibitemShut {NoStop}%
\bibitem [{\citenamefont {Dunjko}\ and\ \citenamefont
  {Briegel}(2017)}]{dunjko2017machine}%
  \BibitemOpen
  \bibfield  {author} {\bibinfo {author} {\bibfnamefont {Vedran}\ \bibnamefont
  {Dunjko}}\ and\ \bibinfo {author} {\bibfnamefont {Hans~J.}\ \bibnamefont
  {Briegel}},\ }\href@noop {} {\enquote {\bibinfo {title} {Machine learning \&
  artificial intelligence in the quantum domain},}\ } (\bibinfo {year}
  {2017}),\ \Eprint {http://arxiv.org/abs/1709.02779} {arXiv:1709.02779
  [quant-ph]} \BibitemShut {NoStop}%
\bibitem [{\citenamefont {Biamonte}\ \emph {et~al.}(2017)\citenamefont
  {Biamonte}, \citenamefont {Wittek}, \citenamefont {Pancotti}, \citenamefont
  {Rebentrost}, \citenamefont {Wiebe},\ and\ \citenamefont
  {Lloyd}}]{Biamonte_2017}%
  \BibitemOpen
  \bibfield  {author} {\bibinfo {author} {\bibfnamefont {Jacob}\ \bibnamefont
  {Biamonte}}, \bibinfo {author} {\bibfnamefont {Peter}\ \bibnamefont
  {Wittek}}, \bibinfo {author} {\bibfnamefont {Nicola}\ \bibnamefont
  {Pancotti}}, \bibinfo {author} {\bibfnamefont {Patrick}\ \bibnamefont
  {Rebentrost}}, \bibinfo {author} {\bibfnamefont {Nathan}\ \bibnamefont
  {Wiebe}}, \ and\ \bibinfo {author} {\bibfnamefont {Seth}\ \bibnamefont
  {Lloyd}},\ }\bibfield  {title} {\enquote {\bibinfo {title} {Quantum machine
  learning},}\ }\href {\doibase 10.1038/nature23474} {\bibfield  {journal}
  {\bibinfo  {journal} {Nature}\ }\textbf {\bibinfo {volume} {549}},\ \bibinfo
  {pages} {195–202} (\bibinfo {year} {2017})}\BibitemShut {NoStop}%
\bibitem [{\citenamefont {Peruzzo}\ \emph {et~al.}(2014)\citenamefont
  {Peruzzo}, \citenamefont {McClean}, \citenamefont {Shadbolt}, \citenamefont
  {Yung}, \citenamefont {Zhou}, \citenamefont {Love}, \citenamefont
  {Aspuru-Guzik},\ and\ \citenamefont {O'Brien}}]{alanVQE}%
  \BibitemOpen
  \bibfield  {author} {\bibinfo {author} {\bibfnamefont {Alberto}\ \bibnamefont
  {Peruzzo}}, \bibinfo {author} {\bibfnamefont {Jarrod}\ \bibnamefont
  {McClean}}, \bibinfo {author} {\bibfnamefont {Peter}\ \bibnamefont
  {Shadbolt}}, \bibinfo {author} {\bibfnamefont {Man-Hong}\ \bibnamefont
  {Yung}}, \bibinfo {author} {\bibfnamefont {Xiao-Qi}\ \bibnamefont {Zhou}},
  \bibinfo {author} {\bibfnamefont {Peter~J}\ \bibnamefont {Love}}, \bibinfo
  {author} {\bibfnamefont {Al{\'{a}}n}\ \bibnamefont {Aspuru-Guzik}}, \ and\
  \bibinfo {author} {\bibfnamefont {Jeremy~L}\ \bibnamefont {O'Brien}},\
  }\bibfield  {title} {\enquote {\bibinfo {title} {{A variational eigenvalue
  solver on a photonic quantum processor}},}\ }\href
  {https://doi.org/10.1038/ncomms5213 10.1038/ncomms5213
  https://www.nature.com/articles/ncomms5213{\#}supplementary-information}
  {\bibfield  {journal} {\bibinfo  {journal} {Nature Communications}\ }\textbf
  {\bibinfo {volume} {5}},\ \bibinfo {pages} {4213} (\bibinfo {year}
  {2014})}\BibitemShut {NoStop}%
\bibitem [{\citenamefont {Farhi}\ \emph {et~al.}(2014)\citenamefont {Farhi},
  \citenamefont {Goldstone},\ and\ \citenamefont {Gutmann}}]{FarhiQAOA}%
  \BibitemOpen
  \bibfield  {author} {\bibinfo {author} {\bibfnamefont {Edward}\ \bibnamefont
  {Farhi}}, \bibinfo {author} {\bibfnamefont {Jeffrey}\ \bibnamefont
  {Goldstone}}, \ and\ \bibinfo {author} {\bibfnamefont {Sam}\ \bibnamefont
  {Gutmann}},\ }\bibfield  {title} {\enquote {\bibinfo {title} {{A Quantum
  Approximate Optimization Algorithm}},}\ }\href
  {http://arxiv.org/abs/1411.4028} {\  (\bibinfo {year} {2014})},\ \Eprint
  {http://arxiv.org/abs/1411.4028} {arXiv:1411.4028} \BibitemShut {NoStop}%
\bibitem [{\citenamefont {Farhi}\ and\ \citenamefont
  {Harrow}(2016)}]{FarhiQAOAsupremacy}%
  \BibitemOpen
  \bibfield  {author} {\bibinfo {author} {\bibfnamefont {Edward}\ \bibnamefont
  {Farhi}}\ and\ \bibinfo {author} {\bibfnamefont {Aram~W}\ \bibnamefont
  {Harrow}},\ }\bibfield  {title} {\enquote {\bibinfo {title} {{Quantum
  Supremacy through the Quantum Approximate Optimization Algorithm}},}\ }\href
  {http://arxiv.org/abs/1602.07674} {\  (\bibinfo {year} {2016})},\ \Eprint
  {http://arxiv.org/abs/1602.07674} {arXiv:1602.07674} \BibitemShut {NoStop}%
\bibitem [{\citenamefont {Preskill}(2018)}]{Preskill_2018}%
  \BibitemOpen
  \bibfield  {author} {\bibinfo {author} {\bibfnamefont {John}\ \bibnamefont
  {Preskill}},\ }\bibfield  {title} {\enquote {\bibinfo {title} {Quantum
  computing in the nisq era and beyond},}\ }\href {\doibase
  10.22331/q-2018-08-06-79} {\bibfield  {journal} {\bibinfo  {journal}
  {Quantum}\ }\textbf {\bibinfo {volume} {2}},\ \bibinfo {pages} {79} (\bibinfo
  {year} {2018})}\BibitemShut {NoStop}%
\bibitem [{\citenamefont {Brassard}\ \emph {et~al.}(2000)\citenamefont
  {Brassard}, \citenamefont {Hoyer}, \citenamefont {Mosca},\ and\ \citenamefont
  {Tapp}}]{brassard2000quantum}%
  \BibitemOpen
  \bibfield  {author} {\bibinfo {author} {\bibfnamefont {Gilles}\ \bibnamefont
  {Brassard}}, \bibinfo {author} {\bibfnamefont {Peter}\ \bibnamefont {Hoyer}},
  \bibinfo {author} {\bibfnamefont {Michele}\ \bibnamefont {Mosca}}, \ and\
  \bibinfo {author} {\bibfnamefont {Alain}\ \bibnamefont {Tapp}},\ }\href@noop
  {} {\enquote {\bibinfo {title} {Quantum amplitude amplification and
  estimation},}\ } (\bibinfo {year} {2000}),\ \Eprint
  {http://arxiv.org/abs/quant-ph/0005055} {arXiv:quant-ph/0005055 [quant-ph]}
  \BibitemShut {NoStop}%
\bibitem [{\citenamefont {Montanaro}(2015)}]{Montanaro_2015}%
  \BibitemOpen
  \bibfield  {author} {\bibinfo {author} {\bibfnamefont {Ashley}\ \bibnamefont
  {Montanaro}},\ }\bibfield  {title} {\enquote {\bibinfo {title} {Quantum
  speedup of monte carlo methods},}\ }\href {\doibase 10.1098/rspa.2015.0301}
  {\bibfield  {journal} {\bibinfo  {journal} {Proceedings of the Royal Society
  A: Mathematical, Physical and Engineering Sciences}\ }\textbf {\bibinfo
  {volume} {471}},\ \bibinfo {pages} {20150301} (\bibinfo {year}
  {2015})}\BibitemShut {NoStop}%
\bibitem [{\citenamefont {Rebentrost}\ \emph {et~al.}(2018)\citenamefont
  {Rebentrost}, \citenamefont {Gupt},\ and\ \citenamefont
  {Bromley}}]{Rebentrost_2018}%
  \BibitemOpen
  \bibfield  {author} {\bibinfo {author} {\bibfnamefont {Patrick}\ \bibnamefont
  {Rebentrost}}, \bibinfo {author} {\bibfnamefont {Brajesh}\ \bibnamefont
  {Gupt}}, \ and\ \bibinfo {author} {\bibfnamefont {Thomas~R.}\ \bibnamefont
  {Bromley}},\ }\bibfield  {title} {\enquote {\bibinfo {title} {Quantum
  computational finance: Monte carlo pricing of financial derivatives},}\
  }\href {\doibase 10.1103/physreva.98.022321} {\bibfield  {journal} {\bibinfo
  {journal} {Physical Review A}\ }\textbf {\bibinfo {volume} {98}} (\bibinfo
  {year} {2018}),\ 10.1103/physreva.98.022321}\BibitemShut {NoStop}%
\bibitem [{\citenamefont {Stamatopoulos}\ \emph {et~al.}(2019)\citenamefont
  {Stamatopoulos}, \citenamefont {Egger}, \citenamefont {Sun}, \citenamefont
  {Zoufal}, \citenamefont {Iten}, \citenamefont {Shen},\ and\ \citenamefont
  {Woerner}}]{stamatopoulos2019option}%
  \BibitemOpen
  \bibfield  {author} {\bibinfo {author} {\bibfnamefont {Nikitas}\ \bibnamefont
  {Stamatopoulos}}, \bibinfo {author} {\bibfnamefont {Daniel~J.}\ \bibnamefont
  {Egger}}, \bibinfo {author} {\bibfnamefont {Yue}\ \bibnamefont {Sun}},
  \bibinfo {author} {\bibfnamefont {Christa}\ \bibnamefont {Zoufal}}, \bibinfo
  {author} {\bibfnamefont {Raban}\ \bibnamefont {Iten}}, \bibinfo {author}
  {\bibfnamefont {Ning}\ \bibnamefont {Shen}}, \ and\ \bibinfo {author}
  {\bibfnamefont {Stefan}\ \bibnamefont {Woerner}},\ }\href@noop {} {\enquote
  {\bibinfo {title} {Option pricing using quantum computers},}\ } (\bibinfo
  {year} {2019}),\ \Eprint {http://arxiv.org/abs/1905.02666} {arXiv:1905.02666
  [quant-ph]} \BibitemShut {NoStop}%
\bibitem [{\citenamefont {Woerner}\ and\ \citenamefont
  {Egger}(2019)}]{Woerner_2019}%
  \BibitemOpen
  \bibfield  {author} {\bibinfo {author} {\bibfnamefont {Stefan}\ \bibnamefont
  {Woerner}}\ and\ \bibinfo {author} {\bibfnamefont {Daniel~J.}\ \bibnamefont
  {Egger}},\ }\bibfield  {title} {\enquote {\bibinfo {title} {Quantum risk
  analysis},}\ }\href {\doibase 10.1038/s41534-019-0130-6} {\bibfield
  {journal} {\bibinfo  {journal} {npj Quantum Information}\ }\textbf {\bibinfo
  {volume} {5}} (\bibinfo {year} {2019}),\
  10.1038/s41534-019-0130-6}\BibitemShut {NoStop}%
\bibitem [{\citenamefont {Egger}\ \emph {et~al.}(2019)\citenamefont {Egger},
  \citenamefont {Gutiérrez}, \citenamefont {Mestre},\ and\ \citenamefont
  {Woerner}}]{egger2019credit}%
  \BibitemOpen
  \bibfield  {author} {\bibinfo {author} {\bibfnamefont {Daniel~J.}\
  \bibnamefont {Egger}}, \bibinfo {author} {\bibfnamefont {Ricardo~Gacía}\
  \bibnamefont {Gutiérrez}}, \bibinfo {author} {\bibfnamefont {Jordi~Cahué}\
  \bibnamefont {Mestre}}, \ and\ \bibinfo {author} {\bibfnamefont {Stefan}\
  \bibnamefont {Woerner}},\ }\href@noop {} {\enquote {\bibinfo {title} {Credit
  risk analysis using quantum computers},}\ } (\bibinfo {year} {2019}),\
  \Eprint {http://arxiv.org/abs/1907.03044} {arXiv:1907.03044 [quant-ph]}
  \BibitemShut {NoStop}%
\bibitem [{\citenamefont {Suzuki}\ \emph {et~al.}(2020)\citenamefont {Suzuki},
  \citenamefont {Uno}, \citenamefont {Raymond}, \citenamefont {Tanaka},
  \citenamefont {Onodera},\ and\ \citenamefont {Yamamoto}}]{Suzuki_2020}%
  \BibitemOpen
  \bibfield  {author} {\bibinfo {author} {\bibfnamefont {Yohichi}\ \bibnamefont
  {Suzuki}}, \bibinfo {author} {\bibfnamefont {Shumpei}\ \bibnamefont {Uno}},
  \bibinfo {author} {\bibfnamefont {Rudy}\ \bibnamefont {Raymond}}, \bibinfo
  {author} {\bibfnamefont {Tomoki}\ \bibnamefont {Tanaka}}, \bibinfo {author}
  {\bibfnamefont {Tamiya}\ \bibnamefont {Onodera}}, \ and\ \bibinfo {author}
  {\bibfnamefont {Naoki}\ \bibnamefont {Yamamoto}},\ }\bibfield  {title}
  {\enquote {\bibinfo {title} {Amplitude estimation without phase
  estimation},}\ }\href {\doibase 10.1007/s11128-019-2565-2} {\bibfield
  {journal} {\bibinfo  {journal} {Quantum Information Processing}\ }\textbf
  {\bibinfo {volume} {19}} (\bibinfo {year} {2020}),\
  10.1007/s11128-019-2565-2}\BibitemShut {NoStop}%
\bibitem [{\citenamefont {Grinko}\ \emph {et~al.}(2019)\citenamefont {Grinko},
  \citenamefont {Gacon}, \citenamefont {Zoufal},\ and\ \citenamefont
  {Woerner}}]{grinko2019iterative}%
  \BibitemOpen
  \bibfield  {author} {\bibinfo {author} {\bibfnamefont {Dmitry}\ \bibnamefont
  {Grinko}}, \bibinfo {author} {\bibfnamefont {Julien}\ \bibnamefont {Gacon}},
  \bibinfo {author} {\bibfnamefont {Christa}\ \bibnamefont {Zoufal}}, \ and\
  \bibinfo {author} {\bibfnamefont {Stefan}\ \bibnamefont {Woerner}},\
  }\href@noop {} {\enquote {\bibinfo {title} {Iterative quantum amplitude
  estimation},}\ } (\bibinfo {year} {2019}),\ \Eprint
  {http://arxiv.org/abs/1912.05559} {arXiv:1912.05559 [quant-ph]} \BibitemShut
  {NoStop}%
\bibitem [{\citenamefont {Nakaji}(2020)}]{nakaji2020faster}%
  \BibitemOpen
  \bibfield  {author} {\bibinfo {author} {\bibfnamefont {Kouhei}\ \bibnamefont
  {Nakaji}},\ }\href@noop {} {\enquote {\bibinfo {title} {Faster amplitude
  estimation},}\ } (\bibinfo {year} {2020}),\ \Eprint
  {http://arxiv.org/abs/2003.02417} {arXiv:2003.02417 [quant-ph]} \BibitemShut
  {NoStop}%
\bibitem [{\citenamefont {Aaronson}\ and\ \citenamefont
  {Rall}(2020)}]{Aaronson_2020}%
  \BibitemOpen
  \bibfield  {author} {\bibinfo {author} {\bibfnamefont {Scott}\ \bibnamefont
  {Aaronson}}\ and\ \bibinfo {author} {\bibfnamefont {Patrick}\ \bibnamefont
  {Rall}},\ }\bibfield  {title} {\enquote {\bibinfo {title} {Quantum
  approximate counting, simplified},}\ }\href {\doibase
  10.1137/1.9781611976014.5} {\bibfield  {journal} {\bibinfo  {journal}
  {Symposium on Simplicity in Algorithms}\ ,\ \bibinfo {pages} {24–32}}
  (\bibinfo {year} {2020})}\BibitemShut {NoStop}%
\bibitem [{\citenamefont {Wang}\ \emph {et~al.}(2020)\citenamefont {Wang},
  \citenamefont {Koh}, \citenamefont {Johnson},\ and\ \citenamefont
  {Cao}}]{wang2020bayesian}%
  \BibitemOpen
  \bibfield  {author} {\bibinfo {author} {\bibfnamefont {Guoming}\ \bibnamefont
  {Wang}}, \bibinfo {author} {\bibfnamefont {Dax~Enshan}\ \bibnamefont {Koh}},
  \bibinfo {author} {\bibfnamefont {Peter~D.}\ \bibnamefont {Johnson}}, \ and\
  \bibinfo {author} {\bibfnamefont {Yudong}\ \bibnamefont {Cao}},\ }\href@noop
  {} {\enquote {\bibinfo {title} {Bayesian inference with engineered likelihood
  functions for robust amplitude estimation},}\ } (\bibinfo {year} {2020}),\
  \Eprint {http://arxiv.org/abs/2006.09350} {arXiv:2006.09350 [quant-ph]}
  \BibitemShut {NoStop}%
\bibitem [{\citenamefont {Grover}(1996)}]{grover1996fast}%
  \BibitemOpen
  \bibfield  {author} {\bibinfo {author} {\bibfnamefont {Lov~K.}\ \bibnamefont
  {Grover}},\ }\href@noop {} {\enquote {\bibinfo {title} {A fast quantum
  mechanical algorithm for database search},}\ } (\bibinfo {year} {1996}),\
  \Eprint {http://arxiv.org/abs/quant-ph/9605043} {arXiv:quant-ph/9605043
  [quant-ph]} \BibitemShut {NoStop}%
\bibitem [{\citenamefont {Petz}\ and\ \citenamefont
  {Ghinea}(2011)}]{petzQfish}%
  \BibitemOpen
  \bibfield  {author} {\bibinfo {author} {\bibfnamefont {D.}~\bibnamefont
  {Petz}}\ and\ \bibinfo {author} {\bibfnamefont {C.}~\bibnamefont {Ghinea}},\
  }\bibfield  {title} {\enquote {\bibinfo {title} {Introduction to quantum
  fisher information},}\ }\href@noop {} {\bibfield  {journal} {\bibinfo
  {journal} {Quantum Probability and Related Topics}\ } (\bibinfo {year}
  {2011})}\BibitemShut {NoStop}%
\bibitem [{\citenamefont {Nielsen}\ and\ \citenamefont
  {Chuang}(2011)}]{Nielsen-Chuang}%
  \BibitemOpen
  \bibfield  {author} {\bibinfo {author} {\bibfnamefont {Michael~A.}\
  \bibnamefont {Nielsen}}\ and\ \bibinfo {author} {\bibfnamefont {Isaac~L.}\
  \bibnamefont {Chuang}},\ }\href@noop {} {\emph {\bibinfo {title} {Quantum
  Computation and Quantum Information: 10th Anniversary Edition}}},\ \bibinfo
  {edition} {10th}\ ed.\ (\bibinfo  {publisher} {Cambridge University Press},\
  \bibinfo {address} {USA},\ \bibinfo {year} {2011})\BibitemShut {NoStop}%
\bibitem [{\citenamefont {Wiebe}\ and\ \citenamefont
  {Granade}(2016)}]{Wiebe_2016}%
  \BibitemOpen
  \bibfield  {author} {\bibinfo {author} {\bibfnamefont {Nathan}\ \bibnamefont
  {Wiebe}}\ and\ \bibinfo {author} {\bibfnamefont {Chris}\ \bibnamefont
  {Granade}},\ }\bibfield  {title} {\enquote {\bibinfo {title} {Efficient
  bayesian phase estimation},}\ }\href {\doibase
  10.1103/physrevlett.117.010503} {\bibfield  {journal} {\bibinfo  {journal}
  {Physical Review Letters}\ }\textbf {\bibinfo {volume} {117}} (\bibinfo
  {year} {2016}),\ 10.1103/physrevlett.117.010503}\BibitemShut {NoStop}%
\bibitem [{\citenamefont {Knill}\ and\ \citenamefont
  {Laflamme}(1998)}]{KnillLaflammeDQC1}%
  \BibitemOpen
  \bibfield  {author} {\bibinfo {author} {\bibfnamefont {E.}~\bibnamefont
  {Knill}}\ and\ \bibinfo {author} {\bibfnamefont {R.}~\bibnamefont
  {Laflamme}},\ }\bibfield  {title} {\enquote {\bibinfo {title} {Power of one
  bit of quantum information},}\ }\href {\doibase 10.1103/PhysRevLett.81.5672}
  {\bibfield  {journal} {\bibinfo  {journal} {Phys. Rev. Lett.}\ }\textbf
  {\bibinfo {volume} {81}},\ \bibinfo {pages} {5672--5675} (\bibinfo {year}
  {1998})}\BibitemShut {NoStop}%
\bibitem [{\citenamefont {Merkel}\ \emph {et~al.}(2013)\citenamefont {Merkel},
  \citenamefont {Gambetta}, \citenamefont {Smolin}, \citenamefont {Poletto},
  \citenamefont {Córcoles}, \citenamefont {Johnson}, \citenamefont {Ryan},\
  and\ \citenamefont {Steffen}}]{Merkel_2013}%
  \BibitemOpen
  \bibfield  {author} {\bibinfo {author} {\bibfnamefont {Seth~T.}\ \bibnamefont
  {Merkel}}, \bibinfo {author} {\bibfnamefont {Jay~M.}\ \bibnamefont
  {Gambetta}}, \bibinfo {author} {\bibfnamefont {John~A.}\ \bibnamefont
  {Smolin}}, \bibinfo {author} {\bibfnamefont {Stefano}\ \bibnamefont
  {Poletto}}, \bibinfo {author} {\bibfnamefont {Antonio~D.}\ \bibnamefont
  {Córcoles}}, \bibinfo {author} {\bibfnamefont {Blake~R.}\ \bibnamefont
  {Johnson}}, \bibinfo {author} {\bibfnamefont {Colm~A.}\ \bibnamefont {Ryan}},
  \ and\ \bibinfo {author} {\bibfnamefont {Matthias}\ \bibnamefont {Steffen}},\
  }\bibfield  {title} {\enquote {\bibinfo {title} {Self-consistent quantum
  process tomography},}\ }\href {\doibase 10.1103/physreva.87.062119}
  {\bibfield  {journal} {\bibinfo  {journal} {Physical Review A}\ }\textbf
  {\bibinfo {volume} {87}} (\bibinfo {year} {2013}),\
  10.1103/physreva.87.062119}\BibitemShut {NoStop}%
\bibitem [{\citenamefont {Blume-Kohout}\ \emph {et~al.}(2013)\citenamefont
  {Blume-Kohout}, \citenamefont {Gamble}, \citenamefont {Nielsen},
  \citenamefont {Mizrahi}, \citenamefont {Sterk},\ and\ \citenamefont
  {Maunz}}]{blumekohout2013robust}%
  \BibitemOpen
  \bibfield  {author} {\bibinfo {author} {\bibfnamefont {Robin}\ \bibnamefont
  {Blume-Kohout}}, \bibinfo {author} {\bibfnamefont {John~King}\ \bibnamefont
  {Gamble}}, \bibinfo {author} {\bibfnamefont {Erik}\ \bibnamefont {Nielsen}},
  \bibinfo {author} {\bibfnamefont {Jonathan}\ \bibnamefont {Mizrahi}},
  \bibinfo {author} {\bibfnamefont {Jonathan~D.}\ \bibnamefont {Sterk}}, \ and\
  \bibinfo {author} {\bibfnamefont {Peter}\ \bibnamefont {Maunz}},\ }\href@noop
  {} {\enquote {\bibinfo {title} {Robust, self-consistent, closed-form
  tomography of quantum logic gates on a trapped ion qubit},}\ } (\bibinfo
  {year} {2013}),\ \Eprint {http://arxiv.org/abs/1310.4492} {arXiv:1310.4492
  [quant-ph]} \BibitemShut {NoStop}%
\bibitem [{\citenamefont {Tanaka}\ \emph {et~al.}(2019)\citenamefont {Tanaka},
  \citenamefont {Uno}, \citenamefont {Suzuki},\ and\ \citenamefont
  {Raymond}}]{githubrepo}%
  \BibitemOpen
  \bibfield  {author} {\bibinfo {author} {\bibfnamefont {T.}~\bibnamefont
  {Tanaka}}, \bibinfo {author} {\bibfnamefont {S.}~\bibnamefont {Uno}},
  \bibinfo {author} {\bibfnamefont {Y.}~\bibnamefont {Suzuki}}, \ and\ \bibinfo
  {author} {\bibfnamefont {R.}~\bibnamefont {Raymond}},\ }\href@noop {}
  {\enquote {\bibinfo {title} {Amplitude estimation without quantum fourier
  transform and controlled grover operators},}\ }\bibinfo {howpublished}
  {\url{https://github.com/qiskit-community/qiskit-community-tutorials/blob/master/algorithms/SimpleIntegral_AEwoPE.ipynb}}
  (\bibinfo {year} {2019})\BibitemShut {NoStop}%
\end{thebibliography}%

\end{document}